\documentclass[epj]{svjour} 

\usepackage[dvips]{graphicx}
\usepackage{float}
\usepackage{epsfig,subfigure}
\usepackage{graphicx}
\usepackage{longtable}
\usepackage{rotating}
\usepackage{amsmath}
\usepackage{amsfonts}
\usepackage{amssymb}
\usepackage{mathrsfs}
\usepackage{calrsfs}
\usepackage{cite}
\usepackage{bm}

\usepackage{txfonts}
\usepackage{pslatex}

\newcommand{\sqrts}{\sqrt{s}}
\newcommand{\sqrtsnn}{\sqrt{s_{\mbox{\tiny{\it{NN}}}}}}
\def\mean#1{\ensuremath{\left<#1\right>}}

\newcommand{\pp}{{$p\mbox{-}p$}}
\newcommand{\ppbar}{{$p\mbox{-}\bar{p}$}}
\newcommand{\AaAa}{{$A\mbox{-}A$}}
\newcommand{\AuAu}{{$Au\mbox{-}Au$}}
\newcommand{\CuCu}{{$Cu\mbox{-}Cu$}}

\newcommand{\Cent}{C}

\newcommand{\pythia}{{\sc pythia}}
\newcommand{\phojet}{{\sc phojet}}
\newcommand{\herwig}{{\sc herwig}}

\begin{document}

\title{Estimates of hadron azimuthal anisotropy from multiparton interactions in proton-proton collisions at $\sqrts$~=~14~TeV }

\author{D.~d'Enterria$^{a,b}$, G.~Kh.~Eyyubova$^{c,d}$, V.~L.~Korotkikh$^c$, I.~P.~Lokhtin$^c$, 
S.~V.~Petrushanko$^c$, L.~I.~Sarycheva$^c$, A.~M.~Snigirev$^c$}

\institute{
$^a$ Laboratory for Nuclear Science, MIT, Cambridge, MA 02139-4307, USA \\ 
$^b$ Institut de Ci\`encies del Cosmos \& ICREA, Univ. Barcelona, 08028 Barcelona, Catalonia \\ 
$^c$ Skobeltsyn Institute of Nuclear Physics, Moscow State University, RU-119991 Moscow, Russia \\ 
$^d$ Department of Physics, University of Oslo,PB 1048 Blindern, N-0316 Oslo, Norway\\
}

\date{Received: date / Revised version: date}

\abstract{
We estimate the amount of collective ``elliptic flow'' expected at mid-rapidity in proton-proton 
(\pp) collisions at the CERN Large Hadron Collider (LHC), 
assuming that any possible azimuthal anisotropy of the produced hadrons with respect to the 
plane of the reaction follows the same overlap-eccentricity and particle-density scalings
as found in high-energy heavy ion collisions. Using a Glauber eikonal model, we 
compute the \pp\ eccentricities, transverse areas and particle multiplicities for various 
phenomenological parametrisations of the proton spatial density.
For realistic proton transverse profiles, we find integrated elliptic flow 
$v_2$ parameters below 3\% in \pp\ collisions at $\sqrts$~=~14~TeV.
}


\titlerunning{Estimates of collective hadron azimuthal anisotropy in \pp\ collisions at $\sqrts$~=~14~TeV }
\authorrunning{D.~d'Enterria {\it et al.}}
\maketitle

\section{Introduction}
\label{sec:intro}

With increasing energies, hadronic collisions are characterised by a larger and larger number 
of produced particles issuing from the fragmentation of a growing number of 
partons involved in the reaction. In the extreme case of high-energy nucleus- nucleus (\AaAa) 
reactions, the multiplicity of released partons is so large -- e.g. $\mathscr{O}$(1000) at midrapidity 
in central \AuAu\ collisions at RHIC energies ($\sqrtsnn$~=~200~GeV) -- that they interact strongly among 
each other leading to 
``hydrodynamical flow'' behaviours~\cite{whitepapers}. A distinct consequence of such partonic 
expansion effects is an azimuthal anisotropy of the final particles produced, $dN/d\Delta\phi$, 
with respect to the reaction plane\footnote{The reaction plane of a collision 
is spanned by the vector of the impact parameter $\vec{b}$ between the centers of the two colliding objects, 
and the beam direction. Its azimuth is given by $\Phi_{RP}$.}, $\Delta\phi=\phi-\Phi_{RP}$. 
At RHIC energies, the large ``elliptical'' anisotropy measured in the data has provided detailed information on the degree 
of collectivity reached in the early stages of nuclear collisions (see e.g.~\cite{Voloshin:2008dg} 
for a recent comprehensive review).\\

The parton density inside hadrons grows very rapidly with increasing energies, as more and more
gluons share smaller and smaller fractional momenta $x$~$\equiv$~$p_{parton}/p_{hadron}$
(see e.g.~\cite{Frankfurt:2005mc,Gelis:2007kn} and refs. therein). Thus, at high enough energies 
the proton itself can be considered as a {\it dense} and {\it extended} partonic object,
and proton-proton collisions can be viewed very much like collisions of ``light nuclei'' composed of their
constituent gluons. The possibility of having multiple parton interactions (MPI) happening simultaneously 
at different impact parameters in hadronic collisions has been discussed since long in the literature~\cite{mpi-th}.
Signals of multiple parton scatterings have been observed experimentally in $p$-$\bar{p}$ collisions 
at centre-of-mass (c.m.) energies  $\mathscr{O}(1$~TeV$)$ in charged particle multiplicity distributions~\cite{Alexopoulos:1998bi} 
as well as in multi-jets events~\cite{Abe:1993rv,Abe:1997xk,Abe:1997bp}. At LHC energies, recent developments of 
general-purpose Monte Carlo (MC) event generators such as \pythia~\cite{Sjostrand:2004pf} and \herwig~\cite{herwig_mpi} 
include an impact-parameter-space description of the \pp\ collision to account for MPI effects. 
As a result of the large number of low-$x$ gluons and increasing multi-parton collisions,
the MC predictions for the {\it total} number of produced particles in ``central'' \pp\ collisions 
at $\sqrts$~=~14~TeV, are pretty large -- up to five hundreds hadrons -- similar to those measured 
e.g. in intermediate-size nuclear (\CuCu) reactions at $\sqrtsnn$~=~200~GeV \cite{Veres:2008nq} 
where a significant elliptic flow has been observed in the data~\cite{Adare:2006ti}. 
Given this situation, it does not seem unjustified to contemplate the possibility of having some type 
``collective'' behaviour also in the final-state of proton-proton collisions at LHC energies~\cite{Drescher:2007jn}.\\

In this paper, we explore the possibility of observing genuine collective expansion effects, 
such as an azimuthal anisotropy with respect to the plane of the reaction, in \pp\ collisions at 
$\sqrts$~=~14~TeV. We note that the partons involved in the scatterings of relevance for such 
collective effects have rather moderate virtualities of the order of the average transverse momentum 
expected in a minimum-bias \pp\ collision at the LHC, i.e. $\mathcal{O}$(0.7~GeV). Recent attempts to 
estimate similar effects have been considered in the context of percolation~\cite{Cunqueiro:2008uu}, 
colour-dipole saturation\footnote{Strictly speaking, the origin of elliptic flow in this work is of a different
nature (initial parton distributions) than the final-state collective effects of interest here.}~\cite{Kopeliovich:2008nx} 
and (viscous) hydrodynamics~\cite{Luzum:2009sb} models. 
In our work, we take a more empirical (i.e. less model-dependent) approach and we simply assume that the 
eccentricity and multiplicity dependence of the elliptic flow experimentally observed in high-energy 
heavy ion collisions (Section~\ref{sec:ellipflow}) holds also for proton collisions at the LHC. 
Using a geometrical eikonal approach (Section~\ref{sec:eikonal}), we determine the 
eccentricity of the transverse overlap region in \pp\ collisions for different proposed proton spatial 
densities (Section~\ref{sec:eikonal_pp}), and then we estimate the resulting associated 
azimuthal anisotropy for various impact-parameters (Section~\ref{sec:results}).

\section{Elliptic flow scaling in ultrarelativistic heavy ion collisions}
\label{sec:ellipflow}

One of the main results at RHIC is the observation~\cite{Ackermann:2000tr,Adcox:2002ms} of a large harmonic modulation 
of the azimuthal distribution of the produced hadrons, $dN/d\Delta\phi$, with respect to the reaction-plane
in \AaAa\ collisions with non-zero impact parameter, i.e. with a lens-shaped overlap zone between the colliding nuclei. 
The measured preferential ``in-plane'' emission is consistent with an efficient translation 
of the initial coordinate-space anisotropy into a final ``elliptical'' asymmetry in 
momentum-space through 
rescatterings between the produced partons in the early  stages of the collision~\cite{Kolb:2000sd,Teaney:2001av}.
The strength of the elliptic flow asymmetry in \AaAa\ collisions is quantified via the second Fourier 
coefficient $v_2 \equiv\mean{cos(2\Delta\phi)}$ of the azimuthal decomposition of the single 
inclusive hadron spectrum with respect to the reaction plane~\cite{Ollitrault:1992bk,Voloshin:1994mz}
\begin{equation}
E\frac{d^3N}{d^3p}=\frac{1}{2\pi}\frac{d^2N}{p_T\,dp_T\,dy}\left(1+2\sum_{n=1}^{\infty}v_n\,cos[n(\phi-\Phi_{RP})]\right).
\end{equation}
The large integrated\footnote{The $v_2$ coefficient is a function of rapidity and transverse momentum, 
and as such it is often referred to as {\it differential} flow. Hereafter, we will be interested in {\it integrated} 
flow, namely the value of the $v_2$ coefficient averaged over transverse momentum and rapidity in a given event.}
$v_2\approx$~0.06 measured in \AuAu\ collisions at $\sqrtsnn$~=~200~GeV~\cite{Ackermann:2000tr,Adcox:2002ms} 
is well reproduced by hydrodynamical simulations~\cite{Kolb:2000sd,Teaney:2001av} that 
predict the development of collective motion along the pressure gradient of the system, which is 
larger for the in-plane directions parallel to the smallest dimension of the lens-shaped overlap zone. 
Within this scenario, one naturally expects 
$v_2$ to be directly proportional to (i) the {\it density} of produced particles in the transverse plane, 
as well as to (ii) the original spatial anisotropy of the system quantified by its {\it eccentricity}, 
usually defined as $\varepsilon=\mean{\vary^2-x^2}/\mean{x^2+\vary^2}$ (where $x,\vary$ are 
the transverse dimensions of the ``lens'', and the average is taken over the initial nuclear profile with 
some weight, see later)~\cite{Heiselberg:1998es,Voloshin:1999gs}.\\

Experimental heavy ion data at different c.m. energies with different colliding systems and varying 
centralities~\cite{Alt:2003ab} indeed indicate a proportionality of the $v_2/\varepsilon$ ratio
with the particle multiplicity normalised by the size of the system, i.e. $v_2/\varepsilon \propto (dN/dy)/A_{\perp}$, 
where $dN/dy$ is the total hadron multiplicity\footnote{Hereafter, $dN/dy$ represents the particle multiplicity 
per unit-rapidity {\it at midrapidity}, i.e. $dN/dy|_{y=0}$, but for simplicity we omit the subindex.} $N$ per unit rapidity $y$, 
and the overlap transverse area $A_{\perp}$ is calculated via a Glauber geometrical model  (see e.g.~\cite{Miller:2007ri})
with standard Woods-Saxon distributions for the initial nuclear distributions~\cite{deJager}.
An attempt to parameterise the observed $v_2/\varepsilon$ dependence on particle density based 
on an ``incomplete equilibration'' model has been developed in Refs.~\cite{Bhalerao:2005mm,Drescher:2007cd}
in terms of the {\em Knudsen number} $K = \lambda/R$, where $\lambda$ is the mean free path of the 
interacting partons 
and $R$ a typical length scale (e.g. the radius) of the system. By definition, 
$K^{-1}$ is the mean number of collisions per particle (i.e. the medium {\it opacity}), and the ideal 
hydrodynamics limit corresponds to $K \rightarrow 0$. Since $\lambda=1/(\sigma_{gg}\,\rho)$, where 
$\sigma_{gg}$ is the effective parton-parton (mostly gluon-gluon) cross section, and since
$\rho(\tau)=1/(\tau\,A_{\perp})\cdot dN/dy$ is the (time-dependent) density of the medium, 
for a typical time $\tau=R/c_s$ (where $c_s$ is the medium speed of sound) one can write the 
(inverse of the) Knudsen number as
\begin{equation}
{K^{-1}}={\sigma_{gg} \;c_s \over A_\perp} {dN \over dy} ,
\label{eq:Knudsen} 
\end{equation}
Within this model, the dependence of $v_2/\varepsilon$ on $dN/dy$ in \AaAa\ collisions can be obtained 
with the simple expression~\cite{Drescher:2007cd}:
\begin{equation}
\Big({v_2\over \varepsilon}\Big)=\Big({v_2^{^{hydro}}\over \varepsilon}\Big){1 \over (1+K/K_0)}, 
\label{eq:v2_scaling}
\end{equation}
where $K$ is given by Eq.~(\ref{eq:Knudsen}) and $K_0\approx$~0.7 is obtained from
a transport model calculation~\cite{Gombeaud:2007ub}.
Thus, taking $c_s=1/\sqrt 3$ for the speed of sound of an ideal parton gas, the model has just two free 
parameters to fit to the data: the effective partonic cross section $\sigma_{gg}$ and the elliptic flow 
in the hydrodynamic-limit $v_2^{hydro}$. A good agreement with \AuAu\ and \CuCu\ data 
at RHIC is obtained with $\sigma_{gg}$~=~5.5~mb and $(v_2^{^{\mbox{\tiny hydro}}}/\varepsilon)$~=~0.22~\cite{Drescher:2007cd}
for a spatial density of the colliding nuclei (needed to determine $\varepsilon$ and $A_\perp$)
based on Color Glass Condensate (CGC) initial conditions\footnote{We note that Eq.~(\ref{eq:v2_scaling2}) is a 
``conservative'' estimate  of the magnitude of $v_2/\varepsilon$ since the alternative initial Glauber matter distribution 
-- with fit parameters $\sigma_{gg}$~=~4.3~mb and $(v_2^{^{\mbox{\tiny hydro}}}/\varepsilon)$~=~0.30 --
results in a  8\%--20\% larger $v_2/\varepsilon$ ratio.}~\cite{cgc}. 
Namely, the elliptic flow data can be well reproduced with the following numerical expression 
\begin{eqnarray}
\Big({v_2\over \varepsilon}\Big) & = & \Big({v_2^{^{hydro}}\over \varepsilon}\Big)\left(\frac{1}{K_0\,\sigma_{gg}\,c_s}+\frac{dN}{dy}\frac{1}{A_\perp}\right)^{-1}\cdot {dN \over dy}\frac{1}{A_\perp} \nonumber \\
& = & 0.22 \cdot \left(0.45+\frac{dN}{dy}\frac{1}{A_\perp\mbox{[mb]}}\right)^{-1}\cdot {dN \over dy}\frac{1}{A_\perp\mbox{[mb]}}.
\label{eq:v2_scaling2}
\end{eqnarray}

In this paper we take as basic assumption that the parton medium produced in the 
overlap region of \pp\ collisions at LHC energies has similar hydrodynamic properties as 
that in \AaAa\ collisions at RHIC and, thus, that the resulting anisotropic flow
parameter $v_2$ follows the same eccentricity and ``transverse'' multiplicity scalings
given by Eqs.~(\ref{eq:Knudsen}) and~(\ref{eq:v2_scaling}). 
For simplicity we will use the same fit-parameters for $\sigma_{gg}$ and 
$(v_2^{^{hydro}}/\varepsilon)$ obtained in~\cite{Drescher:2007cd}, 
i.e. we will assume that Eq.~(\ref{eq:v2_scaling2}) holds too for \pp\ at the LHC.
From the eccentricity $\epsilon$, transverse overlap area $A_\perp$ and
hadron multiplicity $dN/dy$ in \pp\ collisions, determined from a Glauber eikonal model 
using different phenomenological parametrisations of the proton spatial 
density distribution, 
we can thus estimate the expected elliptic flow parameter, $v_2$,
as a function of the \pp\ ``centrality'' or impact parameter $b$.

\section{Eikonal model for proton-proton collisions}
\label{sec:eikonal}

The standard procedure to determine the transverse overlap area, 
eccentricity and final multiplicities in the collision of two nuclei separated by 
impact parameter $b$ is based on a simple Glauber multi-scattering eikonal model 
that assumes straight-line trajectories of the colliding nucleus constituents. 
A recent review that describes the basic formalism can be found in~\cite{Miller:2007ri}.
Often, experimentally it is more useful to introduce the ``reaction centrality'' $\Cent$ as a proxy
for the impact-parameter $b$ of a given collision, by dividing the particle production 
cross section into centrality bins $\Cent_k = \Cent_1, \Cent_2, \cdots $ 
according to some fractional interval $\Delta \Cent$ of the total cross section, e.g. 
$\Delta \Cent = 0.0 - 0.1$ represents the 10\% most central collisions. 
A convenient geometrical definition of  centrality is $\Cent = b^2 / (4~R^2)$, as it corresponds 
to percentiles of the total inelastic cross section in the case of two colliding black disks 
(or any other distribution with small diffuseness) with the same radii $R$, 
see Appendix I for details. 

\paragraph{Thickness and overlap functions, number of binary parton-parton collisions:}
The basic quantity of interest in a Glauber approach is the {\it thickness} (or {\it profile}) {\it function} of the collision
which in the case of a nuclear reaction gives the density of nucleons $\rho$ per unit area $dx\,d\vary$ along the direction 
$z$ separated from the center of the nucleus by an impact parameter $b$, i.e.
$T_A(x,\vary)=A \int dz \, \rho(x,\vary;z)$ for a nucleus of mass number $A$.
By analogy to the nuclear case, the thickness function of a proton with $N_g$ partons
can be written as
\begin{eqnarray}
T_{p}(x,\vary)=N_g \int dz  \; \rho(x,\vary;z),\;\;\\
\mbox{ normalised to }\; \int d^2\vec{b} \,T_{p}(b) = N_g. \nonumber
\label{eq:Tp}
\end{eqnarray}
The {\it overlap function} of a proton-proton collision at $b$ can be obtained as a 
convolution over the corresponding thickness functions of each proton
\begin{eqnarray}
T_{pp}(b)  =  \int dx \, d\vary \; T_{p,1}\left(x+{b/2},\vary\right)\;T_{p,2}\left(x-{b/2},\vary\right),\;\;\\ 
\mbox{ normalised to }\; \int d^2\vec{b} \,T_{pp}(b) = N_g^2.\nonumber
\label{eq:Tpp} 
\end{eqnarray}
For a given parton-parton cross section $\sigma_{gg}$, we can then define the number of binary 
parton-parton collisions in a \pp\ collision at a given impact parameter $b$ 
\begin{eqnarray}
N_{coll,gg}(x,\vary;b) & =& \sigma_{gg} \; T_{p,1}\left(x+{b/2},\vary\right)T_{p,2}\left(x-{b/2},\vary\right)\;,\nonumber \\
N_{coll,gg}(b) & = & \int dx\,d\vary \; N_{coll,gg}(x,\vary;b)\; =\; \sigma_{gg} \,T_{pp}(b).
\label{eq:Ncoll}
\end{eqnarray}
Finally, the probability density of an inelastic parton-parton interaction at impact parameter $b$ 
can be defined as
\begin{eqnarray}
{d^2P_{gg}^{inel}\over d^2\vec{b}}(b)  = {1-e^{-\sigma_{gg} ~T_{pp}(b)} \over \int d^2 \vec{b}\; \left(1-e^{-\sigma_{gg} ~T_{pp}(b)}\right)}, \;\mbox{ or }\;\; \nonumber \\
{dP_{gg}^{inel}\over d b} (b) =2\pi b \; {1-e^{-\sigma_{gg} ~T_{pp}(b)} \over \int d^2 \vec{b} \;  \left(1-e^{-\sigma_{gg} ~T_{pp}(b)}\right)}, 
\label{eq:dPinel_db} 
\end{eqnarray}
according to Poisson statistics. Note that the denominator -- which sums over all events with at least one 
parton-parton interaction -- is just the total inelastic \pp\ cross section $\sigma_{pp}^{inel}$.

\paragraph{Eccentricity and overlap area:}

The eccentricity of a \pp\ collision at impact parameter $b$ can be obtained from the 
asymmetry ratio between the $x$ and $\vary$ ``semi-axis'' dimensions of the overlap zone, 
weighted by the number of parton-parton collisions at $b$:
\begin{equation}
\varepsilon(b)= {\mean{\vary^2 - x^2} \over \mean{\vary^2 +x^2}}=
{ \int dx\,d\vary \; (\vary^2-x^2) \; N_{coll,gg}(x,\vary;b) \over \int dx\,d\vary \; (\vary^2+x^2) \; N_{coll,gg}(x,\vary;b)},
\label{eq:eccent}
\end{equation}
Other weights are possible too, see~\cite{Voloshin:2008dg}, but we take $N_{coll,gg}(b)$ as a natural choice 
-- also used in more sophisticated hydrodynamics approaches~\cite{Luzum:2008cw} -- since our picture for LHC 
energies is based on parton-parton collisions\footnote{Our eccentricity definition, Eq.~(\ref{eq:eccent}), and the 
incomplete thermalisation model one~\cite{Drescher:2007cd} are mathematically identical (modulo, a sign) in the case  
$\mean{x} = \mean{\vary} = \mean{x\,\vary} = 0$, since our spatial densities depend only on $r = \sqrt{x^2 + \vary^2}$ 
and the $N_{coll,gg}(x,\vary)$ weight is an even function of $x$ and $y$.}. As a cross-check, we tested also a weight based on the number 
of participating gluons, obtained from $T_p(x,\vary)$ as described in~\cite{Kolb:2003dz}, obtaining similar results 
for $\varepsilon(b)$ for various proton densities.\\

\noindent
The effective transverse overlap area between the two protons is defined as in~\cite{Drescher:2007cd}:
\begin{equation}
A_{\perp}(b)=4\pi \sqrt {\mean{x^2}}\sqrt {\mean{\vary^2}},
\label{eq:A_perp} 
\end{equation}
where the weighted averages are the same as in Eq.~(\ref{eq:eccent}). We note that there is no commonly accepted 
definition of the absolute normalisation of the overlap area. 
Our area definition with maximum magnitude $4\pi$, is four times larger than that defined in~\cite{Voloshin:1994mz} 
but coincides practically with the geometrical overlap area of two disks with uniform two-dimensional distribution of density. 

\paragraph{Hadron multiplicity:} 
In heavy ion collisions, the centrality dependence of the final hadron multiplicity density is found to 
depend on a combination of the number of nucleon binary collisions and ``participant'' pairs~\cite{whitepapers}. 
Since, at LHC energies, fragmentation of {\it mini}-jets produced in {\it semi}-hard parton-parton scatterings is the largest 
contribution to midrapidity particle production, we will consider 
as a plausible 
operational assumption that the final particle multiplicity in a proton-proton collision 
follows the same impact-parameter-dependence as that of the number of binary parton-parton collisions $N_{coll,gg}(b)$, i.e. that
\begin{equation}
{dN\over dy}(b)= {dN_0 \over dy} \cdot N_{coll,gg}(b) \;, 
\label{eq:dNdy_vs_b} 
\end{equation}
where the absolute normalisation 
$dN_{\mbox{\tiny{\it 0}}}/dy$ is chosen so as to reproduce the {\it minimum-bias} \pp\ multiplicity, 
namely the average multiplicity in a proton-proton collision integrated over all impact-parameters
\begin{equation}
\frac{dN_{\mbox{\tiny{\it MB}}} }{dy}= {dN_0\over dy}
{\int d^2 \vec{b} \; N_{coll,gg}(b) \; {d^2P_{gg}^{inel}\over d^2\vec{b}}(b)}\;, 
\label{eq:dNMB.dyb} 
\end{equation}
which, at midrapidity at LHC energies, is expected to be \\$dN_{\mbox{\tiny{\it MB}}}/dy\approx$~10 according 
to the \pythia~\cite{Sjostrand:2004pf} or \phojet~\cite{phojet} MCs based on various extrapolations from lower energy data.


\subsection{Proton spatial transverse densities}
\label{subsec:p_density}

The matter distribution in transverse space of the colliding protons, $\rho(x,\vary)$ determines 
the probability to have a given number of parton-parton interactions at impact parameter $b$. 
In order to quantify the proton profile function, Eq.~(\ref{eq:Tp}), one often assumes a spherically
symmetric distribution of matter in the Breit system, $\rho(x) d^3x = \rho(r) d^3x$,
and, for simplicity, one takes the same spatial distribution for all parton species in the proton 
(valence and sea quarks,  gluons) and momenta, and neglects any correlations among them.
The spatial proton density has been parameterised with different distributions in the literature 
(see e.g.~\cite{Abe:1997xk,Sjostrand:2004pf}) such as: 
\begin{itemize} 
\item {\it Hard sphere}: The simplest model is to consider that the proton has a spherical form with 
uniform parton density of radius $R$: 
\begin{equation}
\rho(r)={1 \over 4/3\;\pi R^3}\Theta\left(r-R\right)
\label{eq:hardsphere}
\end{equation}
The associated root-mean-square ($rms$) charge radius is \\ $R_{rms}~=~\sqrt{\mean{r^2}}~=~\sqrt{3/5}\,R$.
Of course this is an unrealistic approximation of the proton shape\footnote{E.g. in particular at higher virtualities, such a 
profile violently disagrees with the small-$t$ (where $t$ is the squared momentum exchanged) 
$J/\psi$ data at HERA~\cite{Frankfurt:2004ti}.}, but we keep it as an ``extreme'' case since the 
overlap area of two such distributions has very large eccentricities.

\item {\it Exponential}: Matter can be distributed in the proton according to its {\it charge form factor}, 
which is well represented by an exponential expression in coordinate space:
\begin{equation}
\rho(r)={1 \over 8\pi R^3}e^{-r/R}\;,
\label{eq:expo}
\end{equation}
reproducing to a large extent the spatial distribution of the {\it valence} quarks,
with $rms$ radius $R_{rms}~= \sqrt{12}\,R$.

\item {\it Fermi distribution}: The standard spatial density for nuclei is the 
Fermi-Dirac (or Woods-Saxon) distribution with radius $R$ and surface thickness $a$
\begin{equation}
\rho(r)={\rho_0 \over e^{(r-R)/ a} +1},
\label{eq:WS}
\end{equation}
where $\rho_0$ is a normalisation constant so that $\int d^3 r \,\rho(r)=1$.
The associated $rms$ radius is $R_{rms}= \sqrt{3/5}R$ in the limit $a/R \to 0$ and 
$R_{rms}=\sqrt{12}a$ for $a/R \to \infty$. For $a/R$~=~0.2, $R_{rms}~\approx~1.07\,R$.

\item {\it Gaussian}: A simple Gaussian ansatz, although not very realistic, makes the subsequent 
calculations especially transparent and, therefore, is often used in the literature:
\begin{equation}
\rho(r) = \frac{1}{(\sqrt{2\pi}\,R)^3}\, \exp \left\{- \frac{r^2}{2R^2} \right\} ~.
\label{eq:Gauss}
\end{equation}
The corresponding $rms$ radius is given by $R_{rms}~= \sqrt{3}\,R$.

\item {\it Double-Gaussian}: This corresponds to a distribution with a small core region 
of radius $a_2$ containing a fraction $\beta$ of the total hadronic matter, 
embedded in a larger region of radius $a_1$:
\begin{equation}
\rho(r) \propto \frac{1 - \beta}{a_1^3} \, \exp \left\{
- \frac{r^2}{a_1^2} \right\} + \frac{\beta}{a_2^3} \,
\exp \left\{ - \frac{r^2}{a_2^2} \right\} ~.
\label{eq:doubleGauss}
\end{equation}
\end{itemize}
Although the Gaussian and double-Gaussian are popular proton density choices (e.g. 
in the \pythia\ MC), we will not consider them hereafter because mathematically the convolution of
two such distributions results respectively in an exactly null or very small eccentricity, 
i.e. they cannot generate any elliptic flow according to our ansatz, Eq.~(\ref{eq:v2_scaling2}).\\


\begin{table*}[htp]
\caption{Top rows: Radius $R$ and diffusivity $a$ parameters for various proton spatial densities, 
Eqs.~(\ref{eq:hardsphere})~--~(\ref{eq:WS}), and effective number of gluons $N_g$ in the proton, considered in this work. 
The three bottom rows show the maximum eccentricity and overlap area and inelastic cross section
derived for \pp\ collisions at $\sqrts$~=~14 TeV from each set of parameters within our Glauber approach. 
The last column shows the corresponding values for typical \AuAu\ collisions at RHIC energies.}
\label{tab:1}
\begin{center}
\renewcommand{\arraystretch}{1.2} 
\begin{tabular}{|l|c|c|c|c||c|}\hline\hline
    	& Hard sphere  & Exponential & Fermi-I &  Fermi-II & Fermi  \AuAu \\
    	& refs.~\cite{Hofst,Diaconu:2007zza} &  ref.~\cite{Abe:1997xk}& ref.~\cite{Abe:1997xk}& ref.~\cite{Frankfurt} &ref.~\cite{deJager} \\\hline\hline
radius $R$ (fm)    & 0.89	 & 0.20 & 0.56& 1.05& 6.36	 \\\hline
diffusivity $a$ (fm)   & --	   & 	-- & 0.112 & 0.29  & 0.54	 \\\hline
effective number of partons $N_g$ 	& 17  & 9    & 17   & 4   & 197  {\small (nucleons)} \\\hline \hline 
$rms$ radius $R_{rms}$ (fm)  & 0.69	 & 0.70	 & 0.60 & 1.34 & 5.32 \\ \hline
diffusivity/radius $a/R$    & --	   & 	--  & 0.2 & 0.28 & 0.085	\\\hline 
effective partons/fm$^{3}$ $N_g \rho(r=0\mbox{ fm})$ & 5.8  & 44.8  & 16.7   & 0.48   & 0.17 {\small (nucleons/fm$^3$)}\\\hline\hline 
\pp\ eccentricity $\varepsilon_{max}$ at $b\approx~2\,R_{rms}$ & 1. & -0.4 & 0.07 & -0.05 & 0.4  \\\hline
\pp\ overlap area $A_{\perp}$ at $b$~=~0~fm (fm$^2$)  & 1.6 & 0.58 & 0.85 & 3.8 & 85. \\\hline 
\pp\ inelastic cross section $\sigma_{pp}^{inel}$ (mb) & 80.   & 79.6 & 78.5 & 90. & 7110  \\ \hline \hline 
\end{tabular}
\end{center}
\end{table*}


Table~\ref{tab:1} lists the input parameters used for three of the five proton densities mentioned above. 
The chosen radius $R$ of the hard-sphere parametrisation is consistent with electron-proton scattering 
fits~\cite{Hofst} 
and with diffractive results at HERA 
(more sensitive to the {\it gluon} density)~\cite{Diaconu:2007zza}.
The parameters of the exponential ($R$) and Fermi-I ($R$ and $a$) distributions are those obtained by the 
CDF collaboration in their data analysis 
of double parton scatterings in \ppbar\ collisions at $\sqrts = 1.8$~TeV~\cite{Abe:1997xk}. 
In addition, we quote the parameters of the proton Fermi-II distribution obtained from a 
Fourier transformation of the energy-dependent \pp\ elastic amplitude as discussed in~\cite{Frankfurt}. 
Such an approach aims at taking into account the effective growth of the proton size due to the larger 
transverse spread (``diffusion'') of partons in the proton for increasing collision energy. Within 
this framework, the radius of the strong interactions is expected to be a factor 1.5 larger at the LHC 
compared to fixed-target energies.
From unitarity (i.e. the optical theorem), 
the squared c.m. energy ($s$) dependence of the inelastic (total minus elastic) \pp\ cross section can be 
written as an integral over impact parameters~\cite{Block}:
\begin{equation}
\sigma_{pp}^{inel}(s)=\sigma_{pp}^{tot}(s) - \sigma_{pp}^{el}(s) \\ = \int d^2 \vec{b} \; [2 \mbox{Re}\; \Gamma_{pp}(s;b)-|\Gamma_{pp}(s;b)|^2],
\label{eq:sigma_inel} 
\end{equation}
where $\Gamma_{pp}(s;b)$ is the profile function of the elastic amplitude and 
the integrand represents the distribution of the cross section for generic inelastic 
collisions (i.e., summed over all inelastic final states) over impact parameters.
The $s$-dependence of the inelastic probability defined as 
\begin{equation}
\frac{d^2P_{gg}^{inel}}{d^2\vec{b}}(b) =P_{pp}^{inel}(s;b)= {2 \mbox{Re}\; \Gamma_{pp}(s;b)-|\Gamma_{pp}(s;b)|^2\over \sigma_{pp}^{inel}(s)}.
\label{eq:Pinel}
\end{equation}
can be fitted to the available \pp\ elastic data at high-energies~\cite{Islam} and extrapolated~\cite{Frankfurt}
to LHC energies\footnote{We note that the parton-parton cross section, $\sigma_{gg}$~=~7.8~mb, obtained from 
the fit of~\cite{Frankfurt} is not far from the $\sigma_{gg}$~=~5.5~mb obtained within the incomplete thermalisation 
model, Eq.~(\ref{eq:v2_scaling2}). Both should be interpreted as {\it effective} semihard partonic cross-sections, larger than the 
standard perturbative gluon-gluon cross-section which at LO is $\sigma_{gg} = 9/2\, \pi \alpha^2_s/\mu^2 \approx$~2~mb 
for $\alpha_s$~=~0.5 at a $p_T$-cutoff of order $\mu$~=~1~GeV.}.
By fitting $P_{gg}^{inel}(s;b)$ to Eq.~(\ref{eq:WS}), we obtain the $R$ and $a$ parameters for the Fermi-II spatial
density quoted in Table~\ref{tab:1}.\\

For each one of the density parametrisations considered, we obtain a proton root-mean square ($rms$) charge radius not far
from that obtained from the world-data on $e$-$p$ elastic scattering at low momentum transfer ($Q <$~4~fm$^{-1}$):  
$R_{rms}~\approx$~0.89~fm~\cite{Sick:2003gm}, except for the Fermi-II case which, as aforementioned,
yields an effective proton radius twice larger at the LHC than observed at low energies. 
For comparison purposes with \pp, we add also in the last column of Table~\ref{tab:1} the results for \AuAu\ collisions 
at $\sqrtsnn$~=~200~GeV with the standard $Au$ Fermi distribution parameters~\cite{deJager}.\\

The parton-number normalisation $N_g$ of the proton thickness function, Eq.~(\ref{eq:Tp}), 
can be obtained by requiring that the proton-proton inelastic cross section obtained from the \pp\ 
overlap function, Eq.~(\ref{eq:Tpp}), in the eikonal approximation
\begin{equation}
\sigma_{pp}^{inel}= \int d^2 \vec{b} \; \left(1-e^{-\sigma_{gg} ~T_{pp}(b)}\right)
\label{eq:sigma_inel2} 
\end{equation}
is in the range expected at LHC energies, $\sigma_{pp}^{inel} \approx 80$~mb~\cite{Sjostrand:2004pf,phojet}, 
for our considered effective parton-parton cross section of $\sigma_{gg}=5.5$ mb. 
The corresponding number of partons normalisation $N_g$ is shown in the Table~\ref{tab:1} for each
proton matter distribution. 
We also tabulate the product $N_g \rho(r=0\;\mbox{fm})$ indicating the parton density in the centre of the proton.
For comparison, we also show the density in the center of  the $Au$ nucleus (with $A$~=~197 nucleons), 
which is equal to the well-known value  $A \,\rho_{A}(r=0\mbox{ fm}) = 0.17\mbox{ nucleon/fm}^{3}$.
The corresponding inelastic \AuAu\  cross section at $\sqrtsnn$~=~200~GeV, $\sigma_{AuAu} =  7110$~mb, 
obtained from Eq.~(\ref{eq:sigma_inel2}) with $\sigma_{_{NN}}$~=~42~mb, is larger as expected than that of two black 
absorptive hard disks, $\sigma_{AuAu}$~=~5080~mb, as the Fermi profile for both nuclei has long tails.


\subsection{Number of binary collisions and parton-parton inelastic probability}

\begin{figure*}[htpb]
\centering
\mbox{
\subfigure{\epsfig{figure=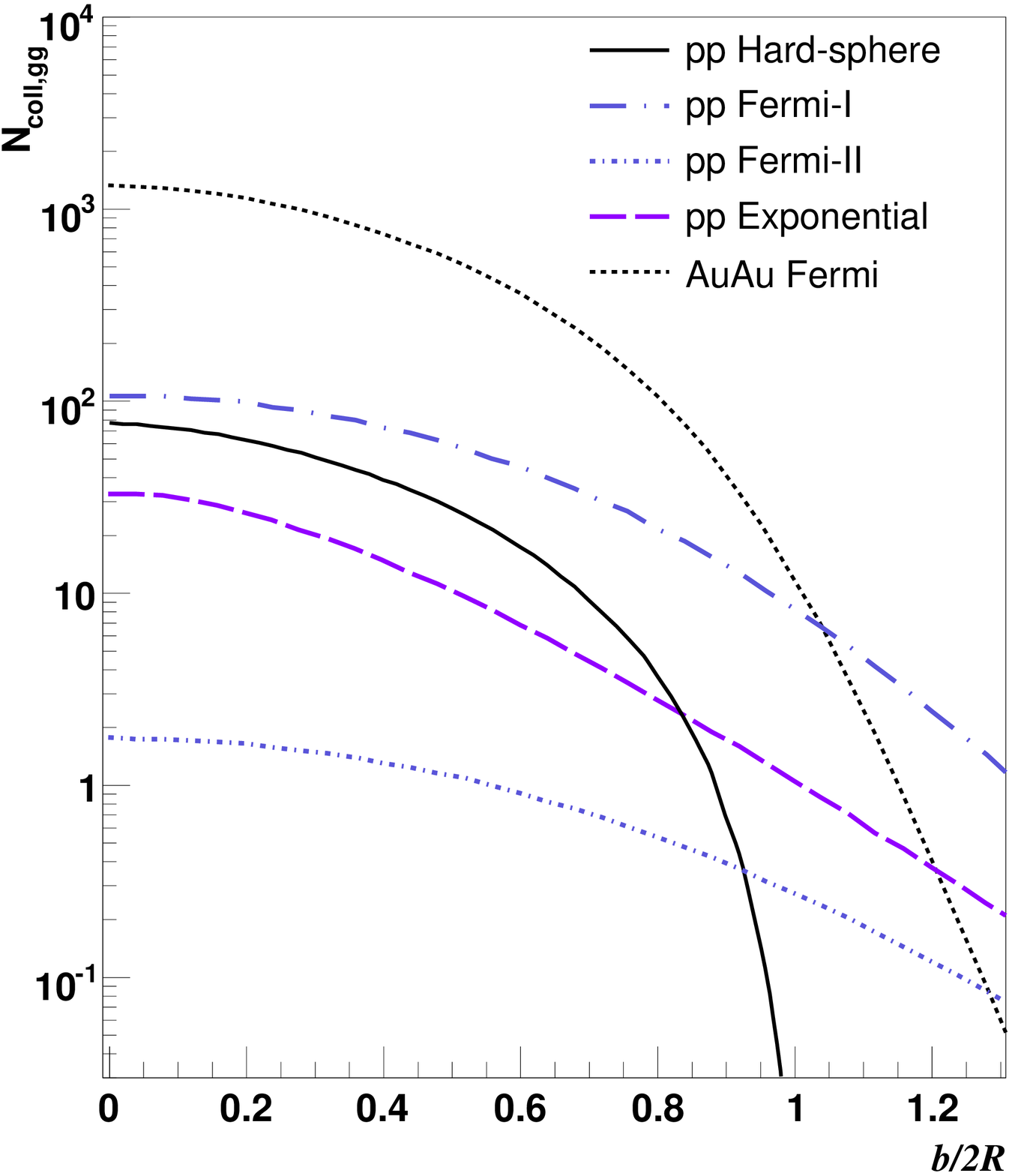,width=8.25cm,height=7.5cm}} 
\subfigure{\epsfig{figure=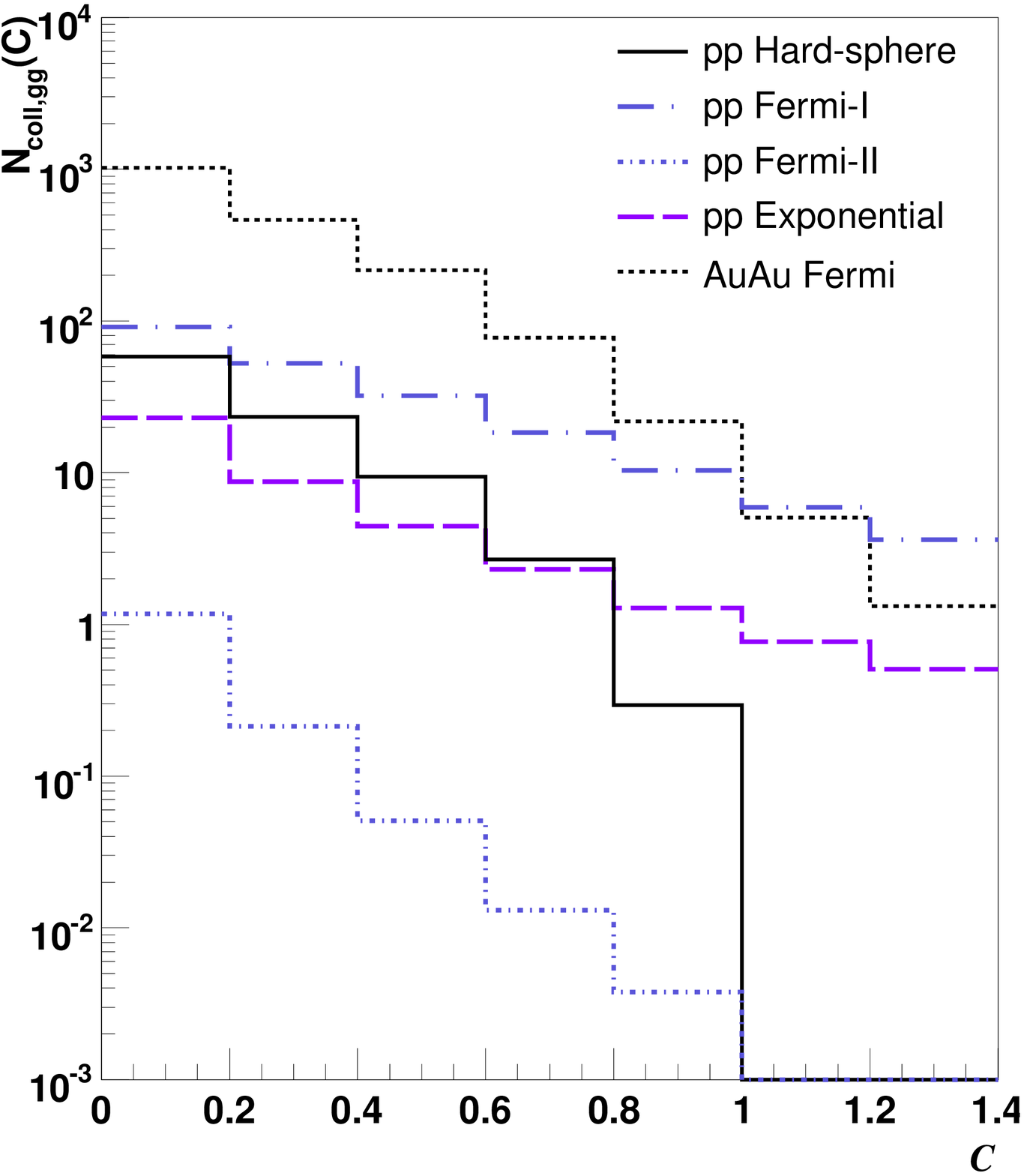,width=8.25cm,height=7.5cm}} 
} 
\caption{
Number of binary parton-parton collisions in \pp\ collisions at $\sqrts$~=~14~TeV 
as a function of scaled impact parameter $b/2R$ (left) and centrality $\Cent$ (right)
for the different proton density distributions considered in this work (Table~\ref{tab:1}).
For comparison, the results for \AuAu\ at RHIC energies 
are shown as a dotted line.}
\label{fig:Ncoll_vs_b}
\end{figure*}

In any eikonal approach, once the spatial densities of the two colliding objects are known, one can derive any geometrical 
quantity related to their interaction from the overlap function, Eq.~(\ref{eq:Tpp}), or equivalently from the number of 
binary collisions of their constituent particles, Eq.~(\ref{eq:Ncoll}).
In the \pp\ case, the number of parton-parton collisions $N_{coll,gg}$ as a function of scaled impact parameter\footnote{For 
the exponential density, for which $R$ and $R_{rms}$ are quite different, we use  $b/(2R_{rms})$.}  $b/2R$ and centrality 
$\Cent = b^2 / (4~R^2)$ for the different proton spatial densities considered in this work are shown in the left and right plots of 
Figure~\ref{fig:Ncoll_vs_b}, respectively. The plot as a function of centrality is shown as an histogram with binning $\Delta \Cent$~=~0.2
as could be obtained experimentally by dividing the data in 20\% fractions of the measured cross section (e.g. 
based on an observable monotonically increasing with impact parameter, such as the event particle multiplicity).\\

\begin{figure*}[htpb]
\centering
\mbox{
\subfigure{\epsfig{figure=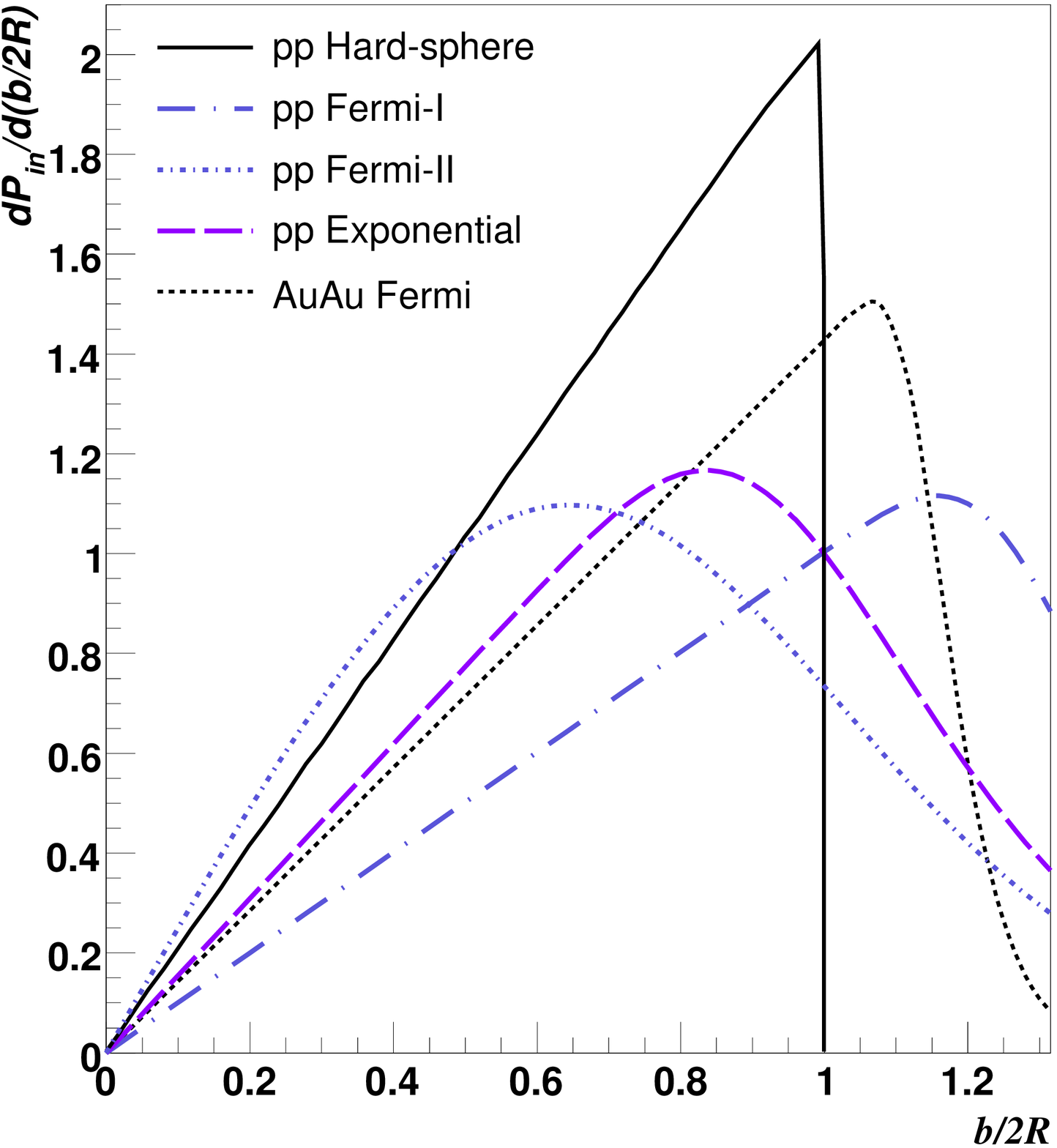,width=8.25cm,height=7.5cm}}
\subfigure{\epsfig{figure=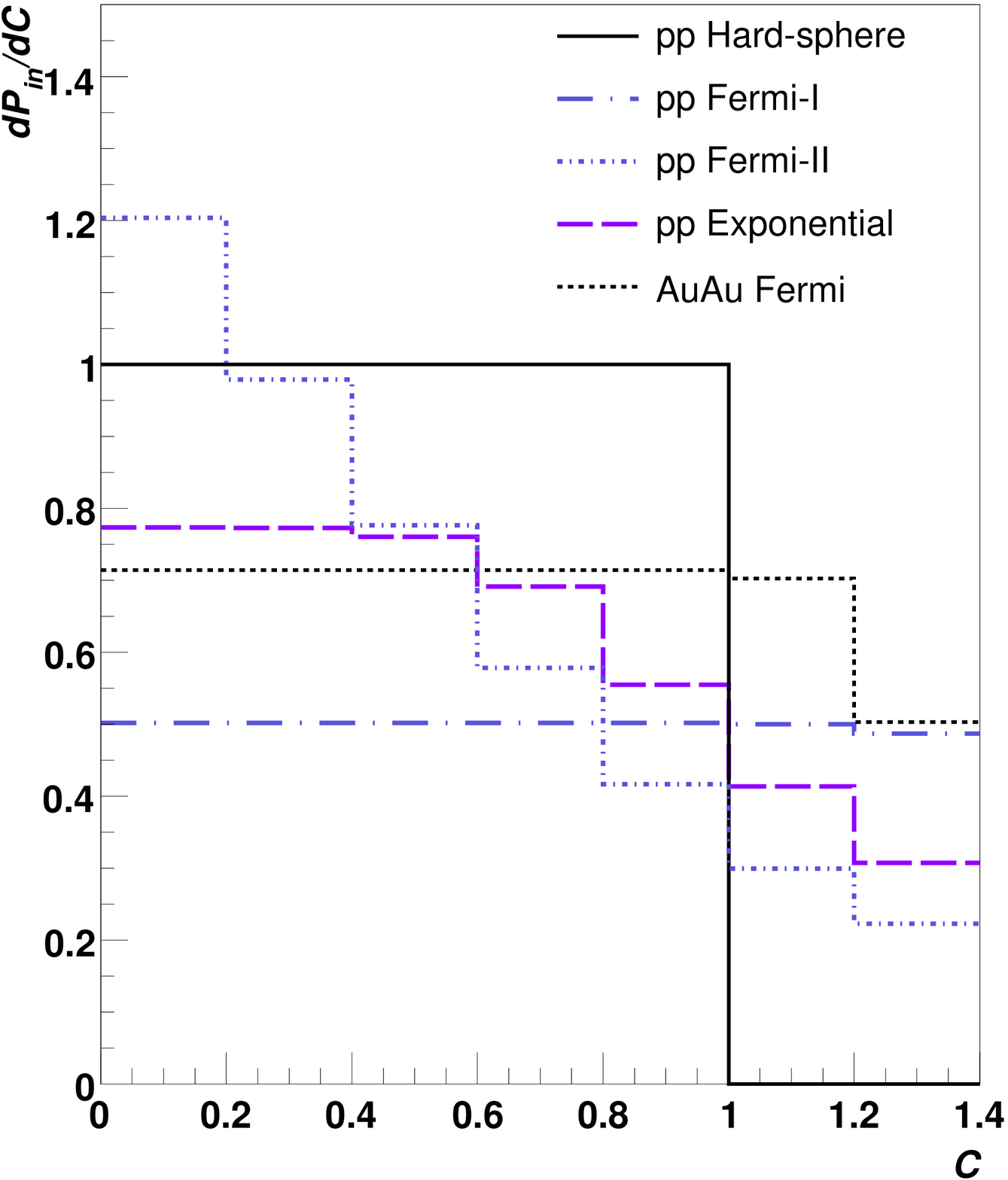,width=8.25cm,height=7.5cm}}
} 
\caption{
Probability density (normalised to unity) of inelastic parton scatterings in \pp\ collisions at $\sqrts$~=~14~TeV
as a function of scaled impact parameter $b/2R$ (left) and centrality $\Cent$ (right)
for the different proton density distributions considered in this work (Table~\ref{tab:1}).
For comparison, the results for \AuAu\ at RHIC energies 
are shown as a dotted line.
\label{fig:dPinel_vs_b}}
\end{figure*}

The exponential and Fermi distributions feature a long tail well beyond twice the radii $R$, 
whereas the hard-sphere parametrisation dives off quickly when $b$ approaches $2\,R$.
For the most central \pp\ collisions ($b$~=~0~fm), the hard-sphere and Fermi-I distributions yield about 
100 parton-parton interactions, the exponential gives about 30 collisions, and the Fermi-II distribution, 
which has a larger proton size and thus a more ``dilute'' parton density, just 2 collisions.
For comparison, the number of nucleon-nucleon collisions in \AuAu\ at RHIC 
shows a steeper centrality-dependence than the Fermi densities for \pp\ collisions because the 
nuclear distribution has a relatively smaller edge diffuseness ($a/R = 0.085$) than the proton one.\\

The other key quantity of our approach is the impact-parame-ter dependence of the midrapidity particle multiplicity 
produced in \pp\ collisions at $\sqrts$~=~14~TeV. Within our framework, we estimate $dN/dy(b)$ via Eqs.~(\ref{eq:dNdy_vs_b})
and (\ref{eq:dNMB.dyb}). The latter depends on the parton-parton inelastic collision probability 
$dP_{gg}^{inel}/db$, Eq.~(\ref{eq:dPinel_db}), 
which is shown as a function of scaled impact-parameter (left) and centrality (right) in Fig.~\ref{fig:dPinel_vs_b} 
for the various proton densities considered in this work. All probabilities are normalised to unity to facilitate the comparison.


\section{Eikonal model results for \pp\ collisions at the LHC}
\label{sec:eikonal_pp}

Based on the eikonal model formalism and proton density parame-trisations discussed in the previous 
section, we derive here all basic quantities needed later to estimate the expected amount of elliptic flow in 
proton-proton collisions at $\sqrts$~=~14~TeV: eccentricity, transverse overlap area and particle multiplicity. 
Again, in order to compare the results for the various proton densities considered in Table~\ref{tab:1} (with 
varying radii $R$), we present our results as a function of the dimensionless {\it scaled} impact-parameter
$b/(2R)$ and centrality $\Cent =b^2 / (4~R^2)$, see Appendix~I.


\paragraph{Eccentricity:}

Figure~\ref{fig:ecc_vs_b} shows the computed eccentricity in \pp\ at $\sqrts$~=~14~TeV 
as a function of the scaled impact-parameter (left) and centrality (right) of the collision.
The eccentricity is increasingly large and positive for the hard-sphere parton density (up to $\varepsilon$~=~1 
for $b~=~2R$) and increasingly negative (down to $\varepsilon = -0.4$ for $b~\approx~2R$) for the 
exponential spatial density. For the Fermi distributions with diffuse edges, the eccentricity is small ($\varepsilon\lesssim$~0.1) 
and changes its sign for very large impact parameters. The larger the ratio of diffuseness-to-radius 
($a/R$~=~0.28, 0.2, and 0.0085 for the Fermi-II, Fermi-I, and Fermi-\AuAu\ respectively) is, the 
smaller the impact-parameter at which $\varepsilon$ changes sign. In any case, the small magnitude of the \pp\ 
eccentricities for the Fermi-I (about five times smaller than Fermi-AuAu) and Fermi-II (very close to zero) profiles 
and the large impact-parameters where the sign change takes place, likely preclude to see any such effect in the elliptic flow data.
Also, as aforementioned, it is easy to show that the overlap of two Gaussian distributions for the 
colliding protons, has zero eccentricity at any impact parameter. 

\begin{figure*}[htpb]
\centering
\subfigure{\epsfig{figure=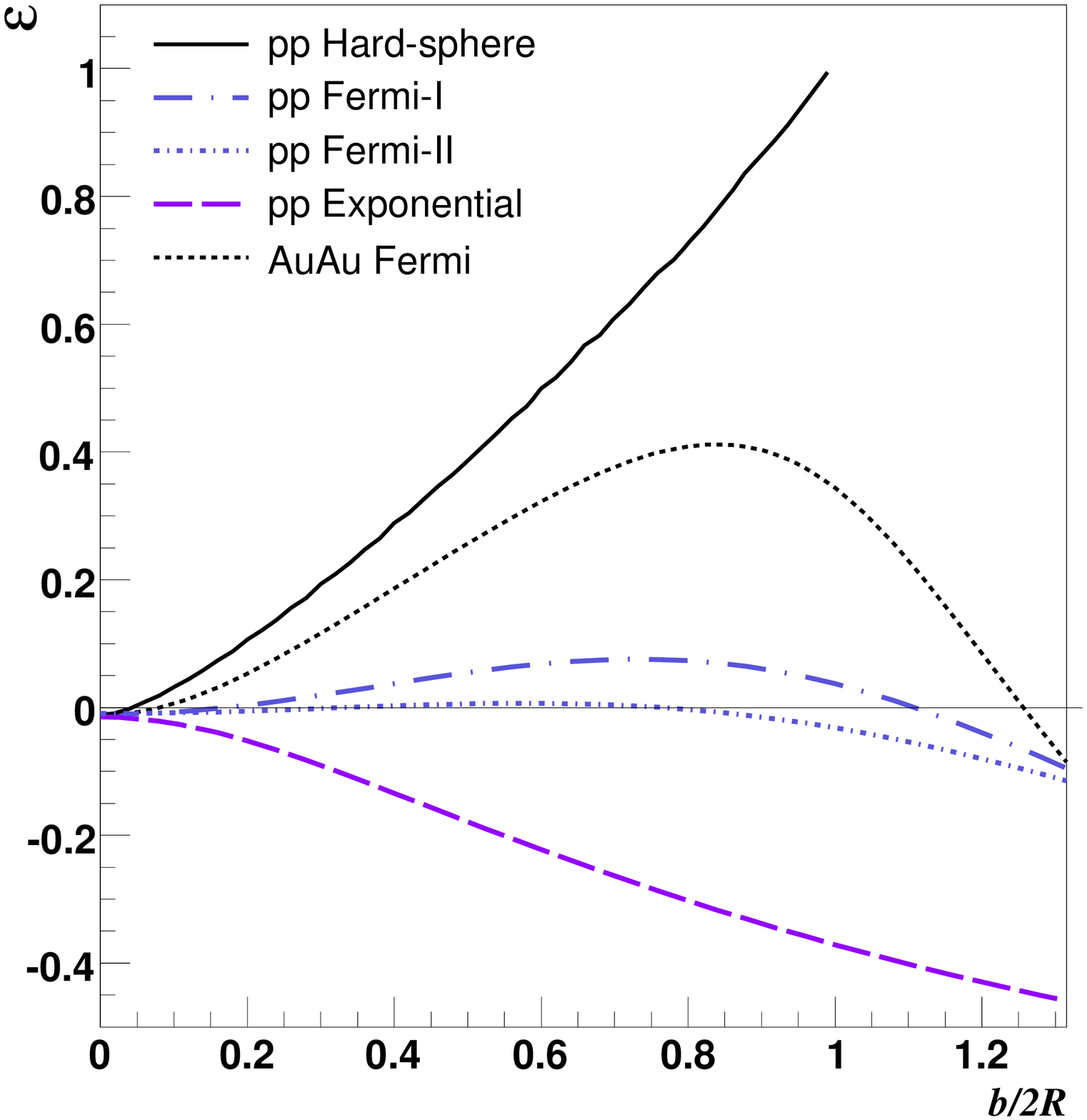,width=8.25cm,height=8.0cm}}
\subfigure{\epsfig{figure=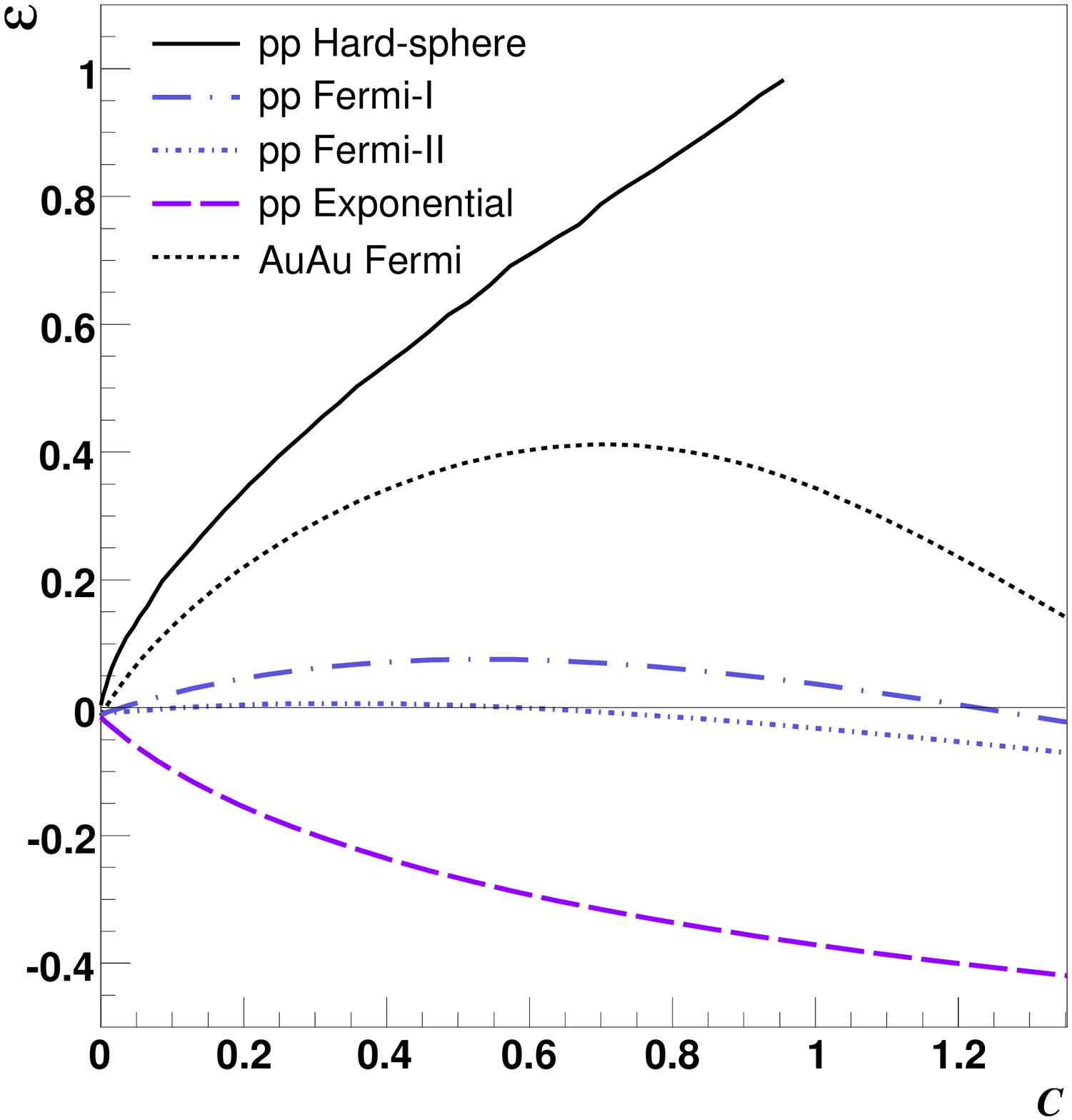,width=8.25cm,height=8.0cm}}
\caption{Eccentricity $\varepsilon$ in \pp\ collisions  at $\sqrts$~=~14~TeV 
as a function of scaled impact parameter $b/2R$ (left) and centrality $\Cent$ (right)
for the different proton density distributions considered in this work (Table~\ref{tab:1}).
For comparison, the results for \AuAu\ at RHIC energies 
are shown as a dotted line.
\label{fig:ecc_vs_b}}
\end{figure*}


\paragraph{Overlap area:}

\begin{figure*}[htpb]
\centering
\subfigure{\epsfig{figure=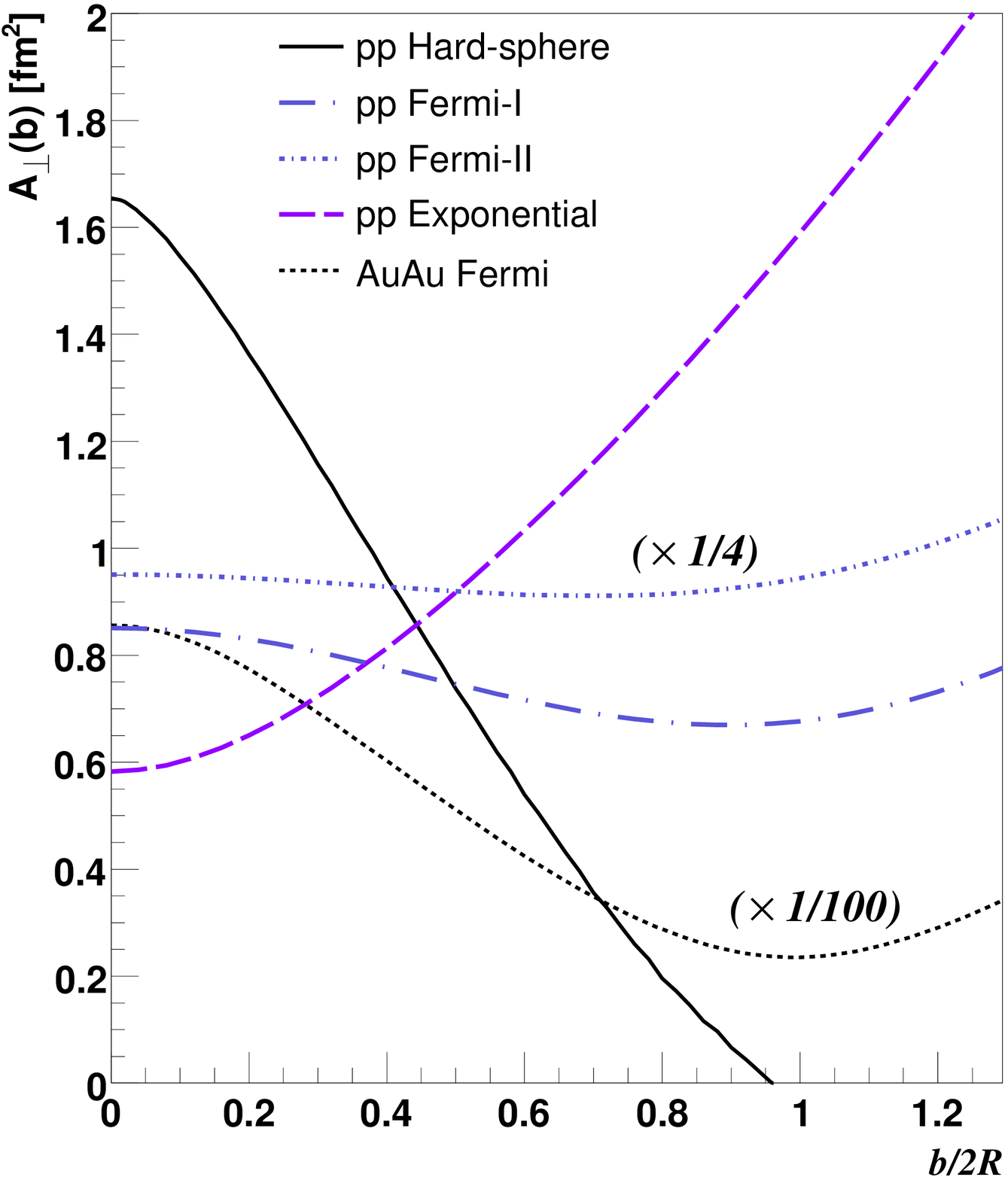,width=8.25cm,height=7.5cm}}
\subfigure{\epsfig{figure=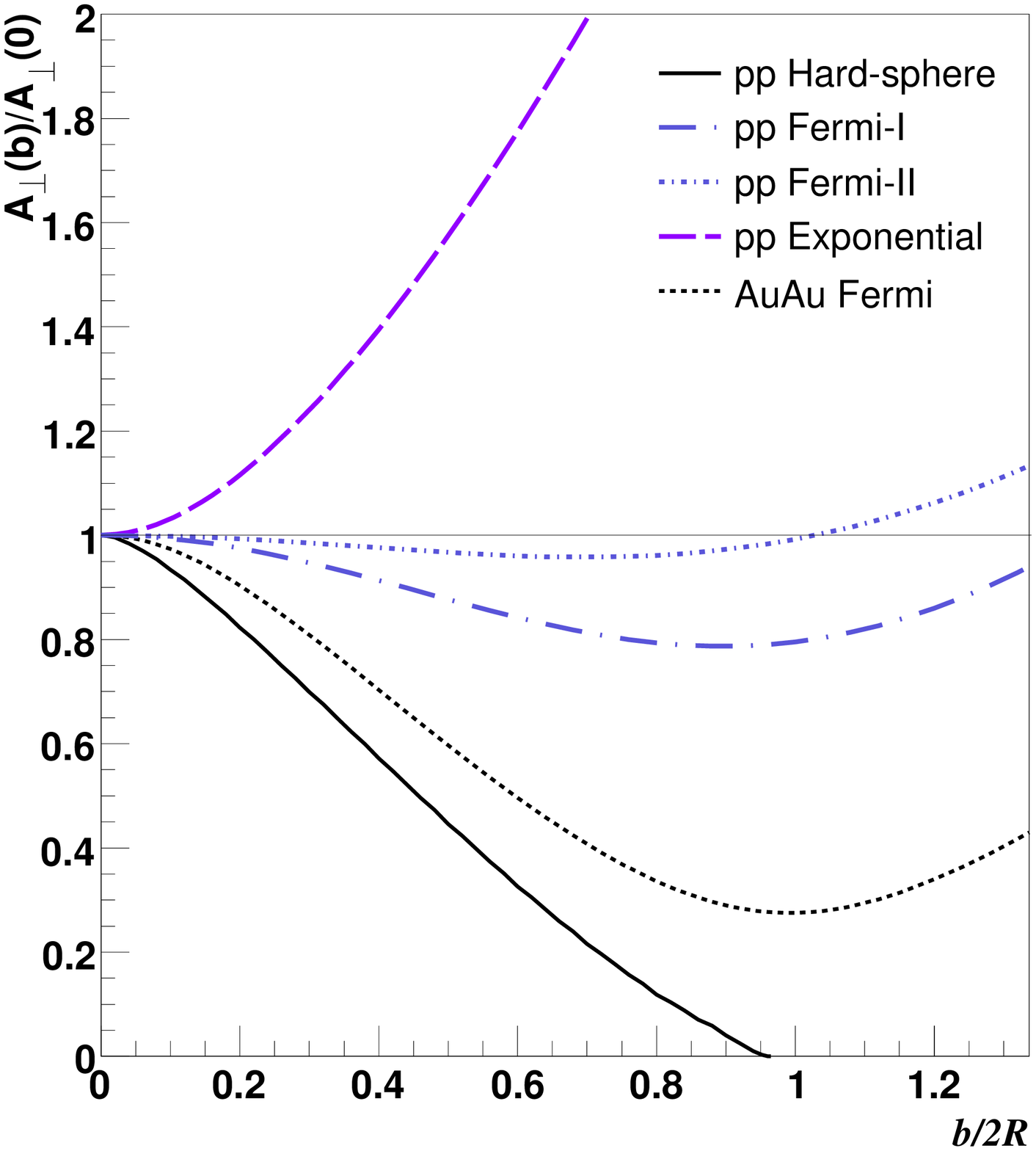,width=8.25cm,height=7.5cm}}
\caption{Effective overlap area $A_\perp$ in \pp\ collisions  at $\sqrts$~=~14~TeV 
as a function of scaled impact parameter $b/2R$ for the different proton density distributions 
considered in this work (Table~\ref{tab:1}). For comparison, the results for \AuAu\ 
at RHIC energies 
are shown as a dotted line. The left plot shows the absolute value of $A_\perp$ 
(for clarity, the Fermi-II and \AuAu\ curves are scaled by factors of 1/4 and 1/100 
respectively). The right plot shows the area normalised to the value for 
central collisions, $A_{\perp}(b)/A_\perp(b=0\mbox{ fm})$.
\label{fig:AT_vs_b}}
\end{figure*}

Figure~\ref{fig:AT_vs_b} shows the \pp\ transverse overlap area as a function of scaled impact parameter. 
The left plot shows the absolute $A_{\perp}(b)$ value, and the right plot presents the overlap-area normalised
to the maximum value for a ``head-on'' collision at zero impact-parameter, $A_{\perp}(b)/A_\perp(b=0\mbox{ fm})$.
The overlap area for central \pp\ collisions is largest for the Fermi-II density ($A_\perp \approx$~4~fm$^{2}$) 
and smallest for the exponential distribution ($A_\perp \approx$~0.6~fm$^{2}$). Again, as for the eccentricity, the hard-sphere and exponential distribution have opposite
behaviours as a function of centrality: $A_{\perp}(b)$ decreases (increases) for the former (for the latter).
On the other hand, there is only a small dependence of the transverse area on impact-parameter for the overlap of two Fermi 
densities with diffuse edges. Compared to \AuAu, the obtained values of $A_{\perp}$ in \pp\ collisions, for the Fermi
and hard-sphere densities ($A_\perp \approx$~1~fm$^{2}$), are about two orders of magnitude smaller.

\paragraph{Hadron multiplicity:}

\begin{figure*}[htpb]
\centering
\mbox{
\subfigure{\epsfig{figure=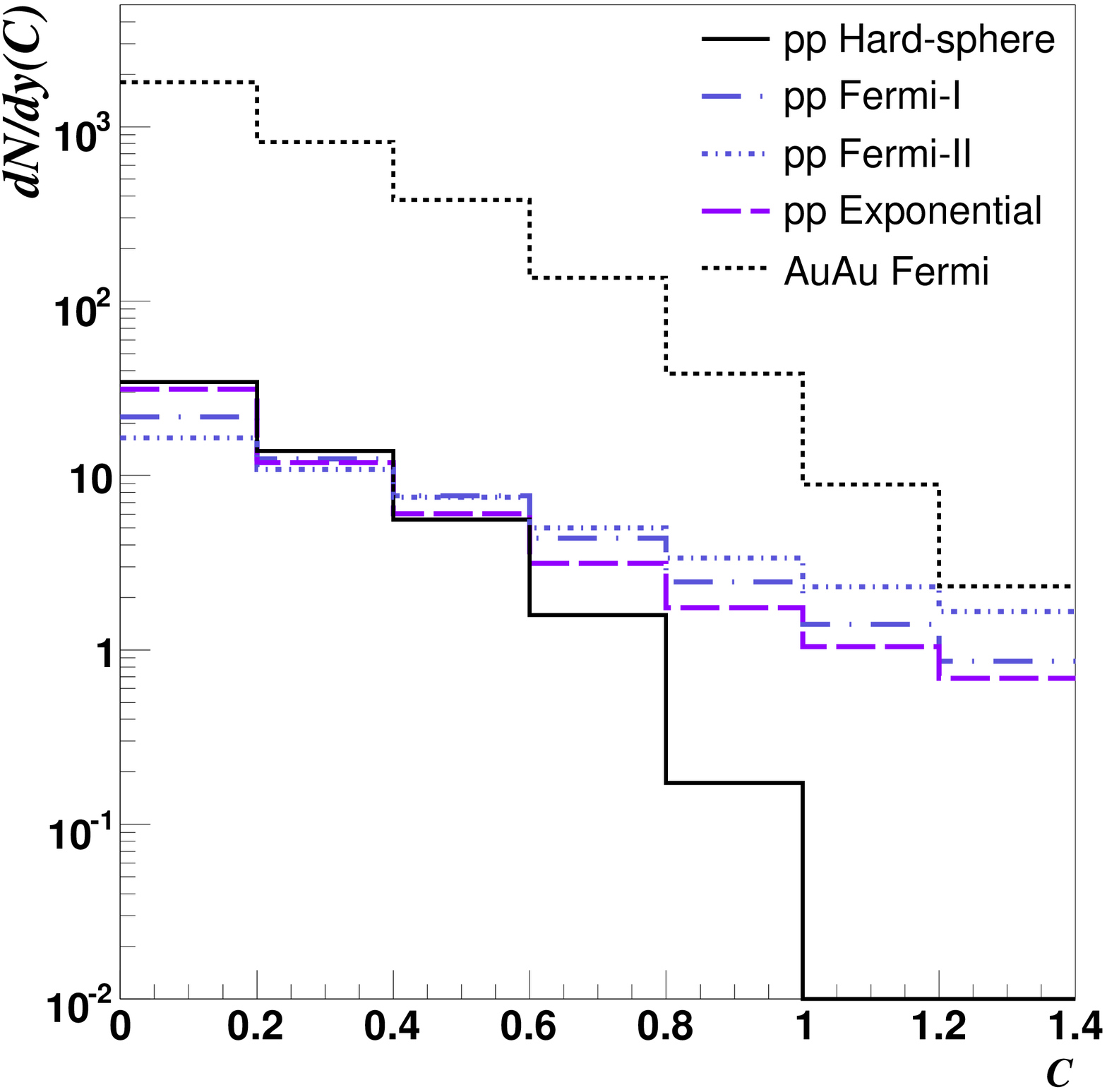,width=8.25cm,height=7.5cm}}
\subfigure{\epsfig{figure=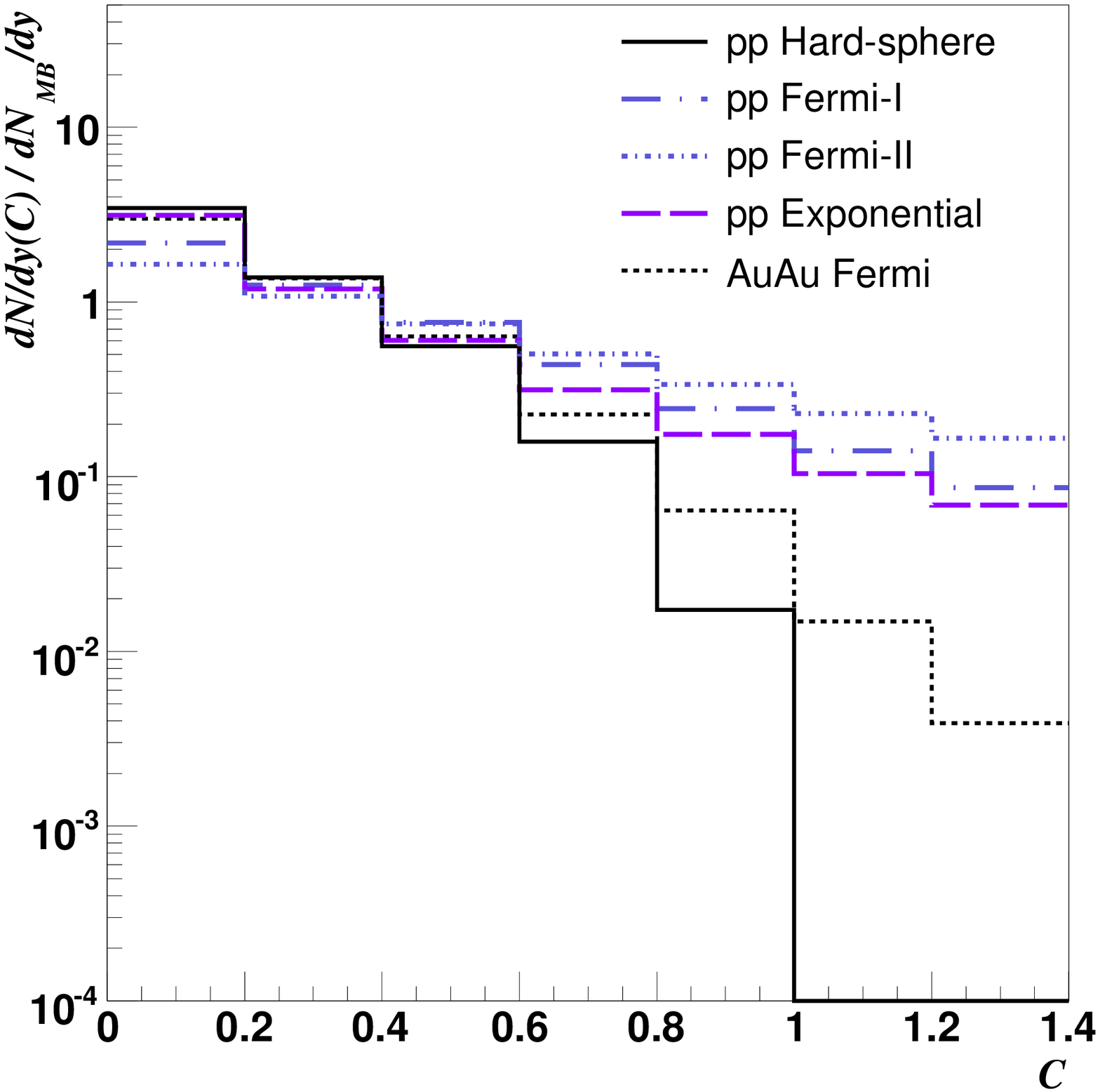,width=8.25cm,height=7.5cm}}
} 
\caption{
Absolute (left) and normalised (right) particle multiplicity at midrapidity as a function of  centrality $\Cent$ 
in \pp\ collisions at $\sqrts$~=~14~TeV for the different proton density distributions considered in this work (Table~\ref{tab:1}).
For comparison, the results for \AuAu\ at RHIC energies 
are shown as a dotted line.
\label{fig:dNdy_vs_b}}
\end{figure*}

The third ingredient of the incomplete thermalisation model, Eq.~(\ref{eq:v2_scaling2}), is the 
particle multiplicity per unit rapidity at midrapidity, $dN/dy$, obtained via Eqs.~(\ref{eq:dNdy_vs_b})~--~(\ref{eq:dNMB.dyb}).
The results are shown in Fig.~\ref{fig:dNdy_vs_b}  for a value\footnote{Note that, given
our absolute normalisation based on $dN_{\mbox{\tiny{\it MB}}}/dy$, the actual $N_g$ and $\sigma_{gg}$ values 
do not play practically any role in the final determination of the hadron multiplicity.} of $dN_{0}/dy$ chosen
in Eq.~(\ref{eq:dNMB.dyb}) so as to reproduce the expected $dN_{\mbox{\tiny{\it MB}}}/dy=10$ 
multiplicity in minimum-bias \pp\ collisions at the LHC. All proton densities yield similar values
of $dN/dy$ in a wide range of centralities $\Cent$ in \pp\ collisions. The 20\% most central collisions 
result in  2--3 times larger multiplicities than in minimum bias \pp\ (or \AuAu) collisions for all considered densities.
For the comparison \AuAu\ data, we take ${dN/dy} = 600$.\\

\begin{figure*}[htpb]
\centering
\mbox{
\subfigure{\epsfig{figure=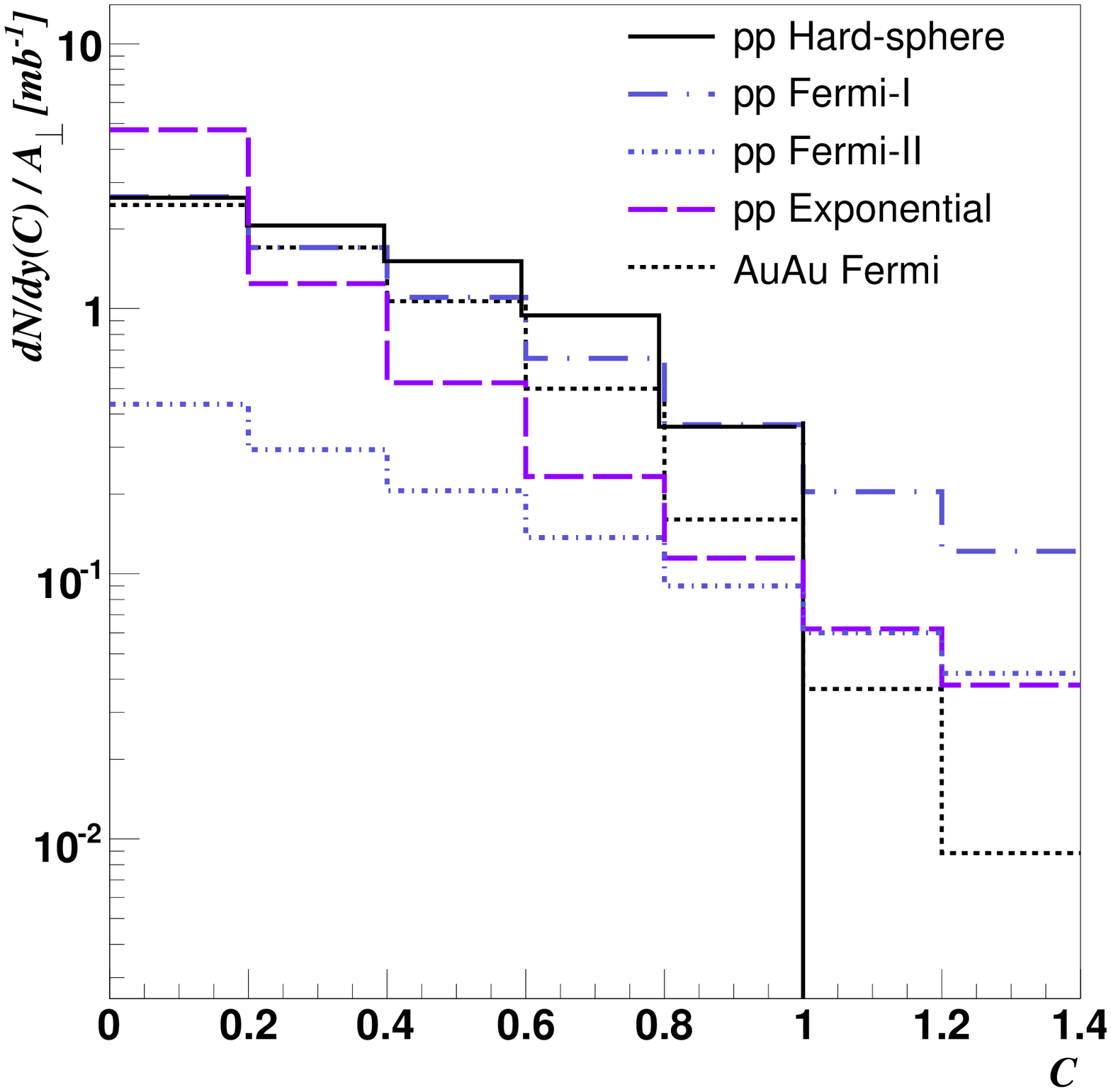,width=9.5cm,height=7.5cm}}
} 
\caption{
Particle multiplicity density at $y=0$ per unit of transverse overlap area $A_{\perp}$ 
in \pp\ collisions at $\sqrts$~=~14~TeV as a function of centrality $\Cent$ 
for the different proton density distributions considered in this work (Table~\ref{tab:1}).
For comparison, the results for \AuAu\ at RHIC energies 
are shown as a dotted line.
\label{fig:dNdy_AT_vs_b}}
\end{figure*}

As we see from our elliptic-flow ``ansatz'', Eq.~(\ref{eq:v2_scaling2}), the $v_2$ parameter 
depends on the particle multiplicity but only after normalisation by the size of the system.
Figure~\ref{fig:dNdy_AT_vs_b} shows the transverse density multiplicity, namely 
the ratio of the hadron multiplicity over the transverse overlap area $(dN/dy)/A_\perp$, 
for different forms of the proton matter distribution. We find ratios $(dN/dy)/A_\perp~\approx$~2.5 -- 4.5~mb$^{-1}$ 
for all distributions  in central \pp\ collisions similar to those found in \AuAu\ collisions at RHIC except for the dilute Fermi-II density,
which features a much lower $(dN/dy)/A_\perp~\approx$~0.5~mb$^{-1}$ value.
Interestingly, although the particle multiplicity is about two orders of magnitude larger in nuclear than 
in proton collisions, the overlap area is roughly smaller by the same amount (see Fig.~\ref{fig:AT_vs_b}) 
and, therefore, the multiplicity density is not very dissimilar in \pp\ and \AaAa\ collisions: 
$(dN/dy)/A_\perp \approx$~0.1~--~4.5~mb$^{-1}$, for all densities within a large interval of centralities.


\section{Elliptic flow estimates in \pp\ collisions at the LHC}
\label{sec:results}

\begin{figure*}[htpb]
\centering
\mbox{
\subfigure{\epsfig{figure=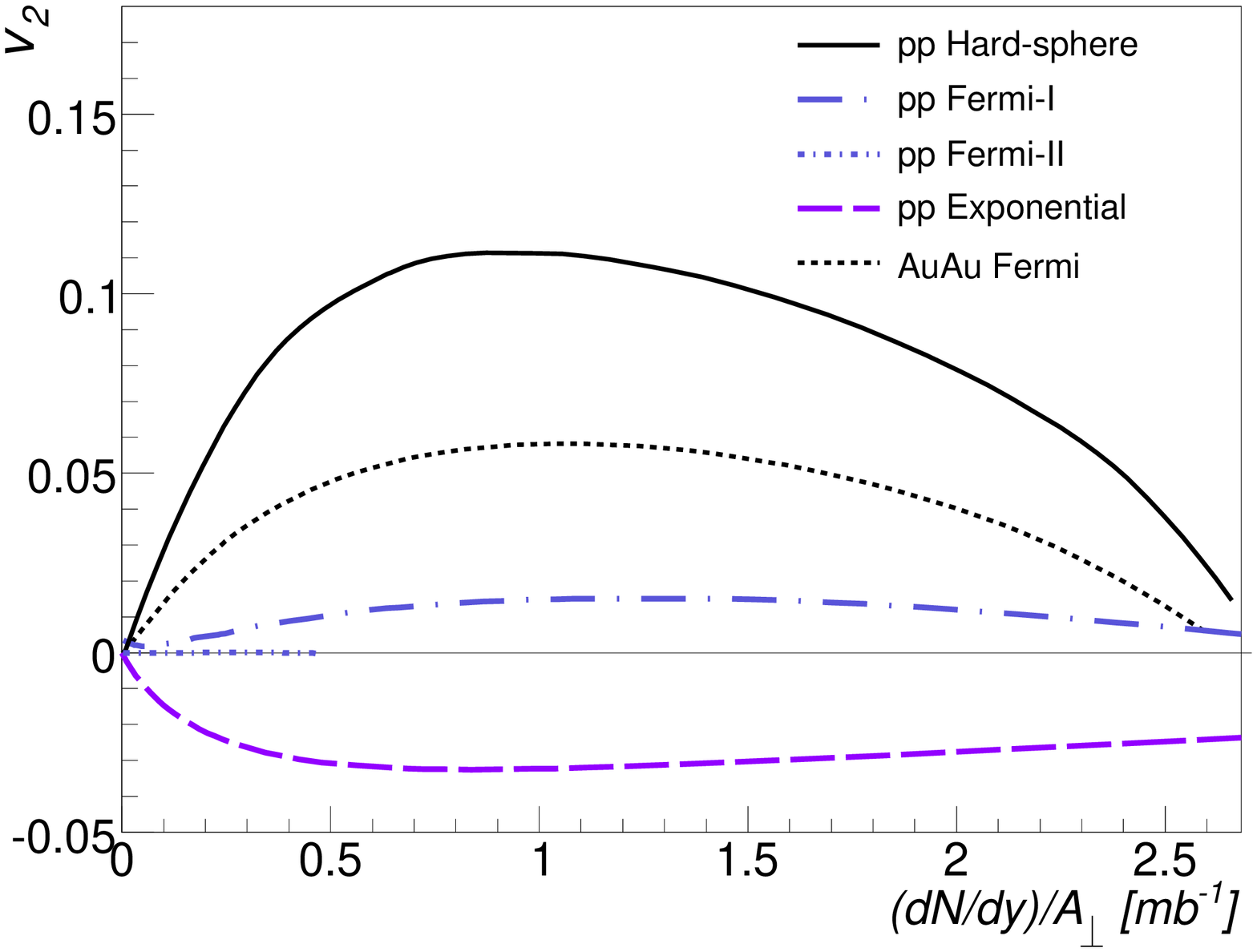,width=11.cm,height=7.5cm}}
}
\caption{
Integrated elliptic flow $v_2$ parameter as function of transverse particle multiplicity, 
$(dN/dy)/A_\perp$, 
expected at midrapidity in \pp\ collisions  at $\sqrts$~=~14~TeV for the different proton density distributions considered 
in this work (Table~\ref{tab:1}). For comparison, the \AuAu\ results are shown as a dotted line.
\label{fig:v2pp}}
\end{figure*}

\begin{figure*}[htpb]
\centering
\mbox{ 
\subfigure{\epsfig{figure=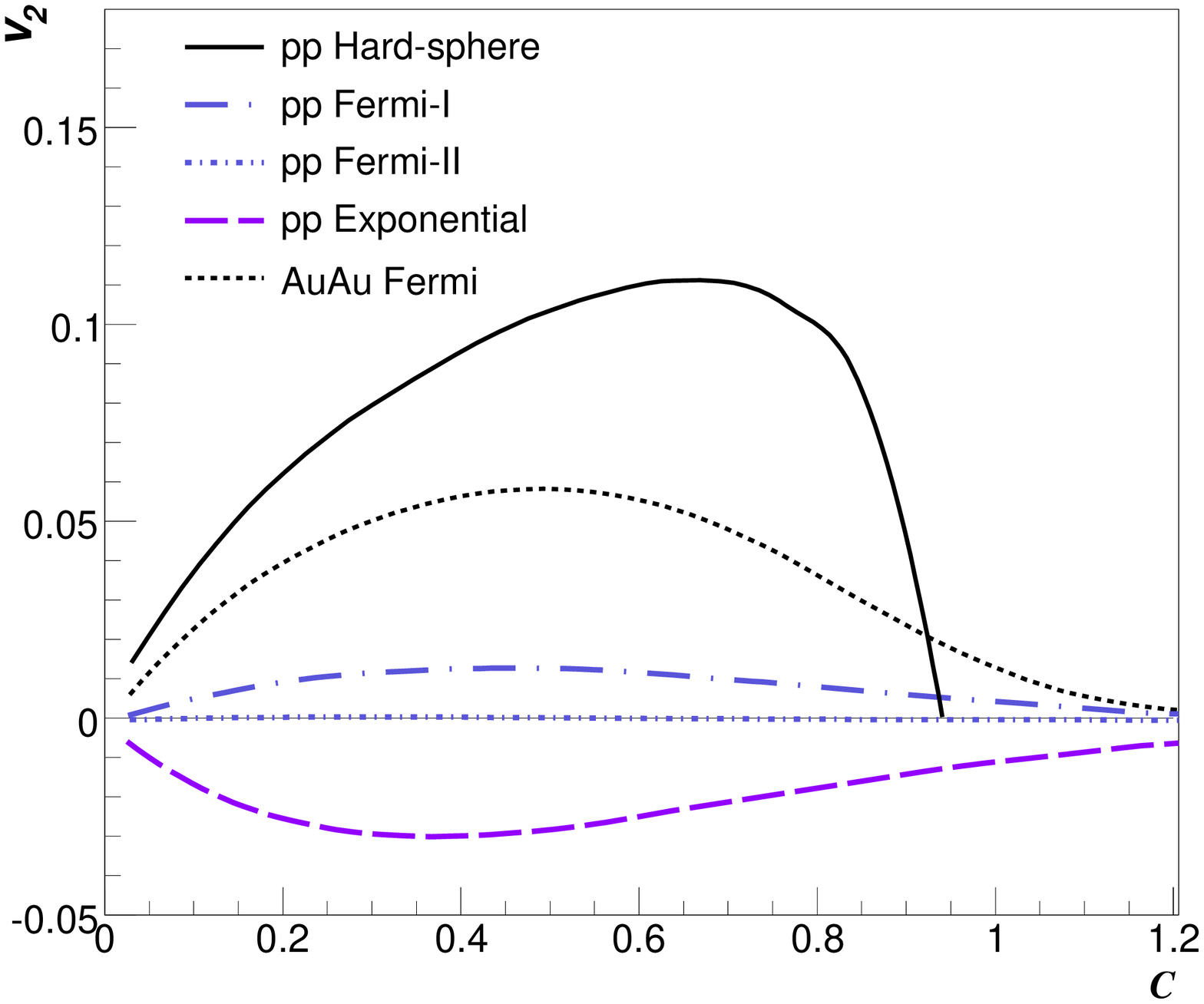,width=8.5cm}}
\subfigure{\epsfig{figure=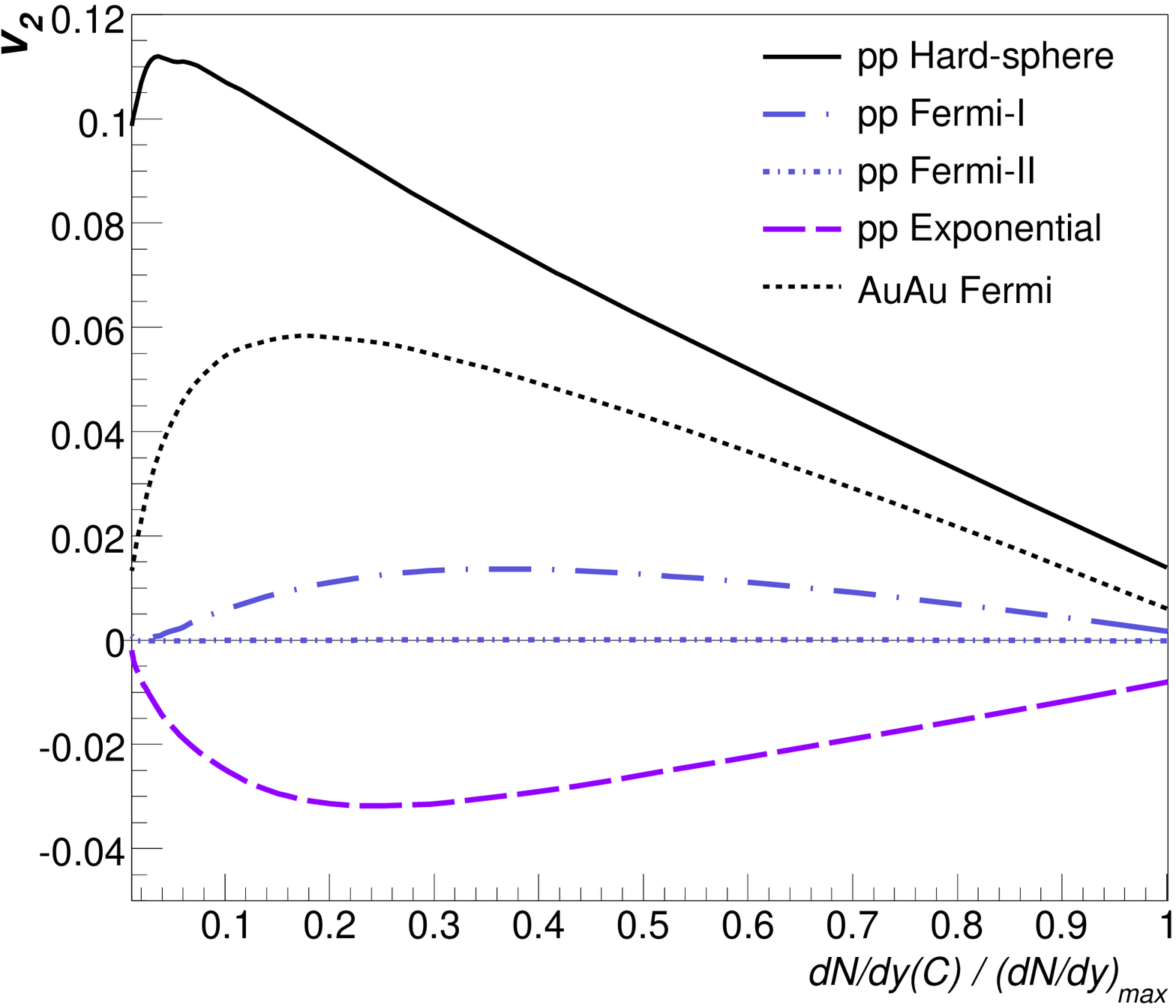,width=8.5cm}}
}
\caption{
Integrated elliptic flow $v_2$ parameter as function of centrality  (left panel) and of normalised particle multiplicity (right panel)
at midrapidity in \pp\ collisions  at $\sqrts$~=~14~TeV for the different proton density distributions considered in this work (Table~\ref{tab:1}).
For comparison, the $v_2$ for \AuAu\ at RHIC energies is shown as a dotted line.
\label{fig:v2pp_vs_cent}}
\end{figure*}

From the values of eccentricity $\varepsilon$ (Fig.~\ref{fig:ecc_vs_b}), overlap area $A_{\perp}$ 
(Fig.~\ref{fig:AT_vs_b}) and hadron multiplicity $dN/dy$ (Fig.~\ref{fig:dNdy_vs_b}), as a function 
of impact-parameter $b$ determined in the previous chapter, and using the Eq.~(\ref{eq:v2_scaling2}) 
of the incomplete thermalisation model, we can finally obtain 
the value of the integrated elliptic flow parameter $v_2$ for each one of the four different proton densities. 
By construction, $v_2(b)$ will follow the general behaviour of $\varepsilon(b)$. 
Figure~\ref{fig:v2pp} shows the integrated $v_2$  as a function of transverse particle density 
expected at midrapidity in \pp\ collisions at LHC energies. 
In Fig.~\ref{fig:v2pp_vs_cent} we show $v_2$ versus centrality (left) and versus 
particle multiplicity (right) normalised by the maximum value in the most central events, 
$N(\Cent)/N_{max}$, which is a proxy for the reaction centrality $\Cent$ often used experimentally.\\

Several features are noticeable in both figures. 
First, the hard-sphere distribution features a maximum value of $v_2~\approx$~0.1, almost
twice higher than in \AuAu\ collisions and much larger than the rest of \pp\ elliptic flow parameters. 
The fact that the overlap of two hard-spheres with infinitely sharp edges (see Fig.~\ref{fig:ecc_vs_b})
yields artificially large eccentricities is a well-known fact~\cite{Voloshin:2008dg}. We plot this result 
just to delimit the theoretically maximum amount of $v_2$ that could be generated within our approach. 
Second, the maximum $v_2$ obtained with the Fermi-I profile is about 1.5\% 
and vanishingly small for the dilute Fermi-II distribution. 
Third, exponential transverse densities for the proton, feature {\it negative} \pp\ eccentricities with minimum 
elliptic flow parameters of the order of $-$3.5\%. A negative integrated $v_2$ would indicate that the anisotropy of hadron 
emission is not {\it in-plane} (as for a positive $v_2$) but {\it out-of-plane}, i.e. perpendicular to the reaction plane.\\

Setting aside the $v_2$ obtained with the unrealistic hard-sphere profile, the maximum absolute values of the integrated elliptic flow in \pp\ collisions 
in the range 1\%--3\% are consistent with those obtained in a recent hydrodynamics study~\cite{Luzum:2009sb}
-- which uses a spatial density based on a parametrisation of the Fourier transform of the proton electromagnetic 
form-factor~\cite{Miller:2007uy} -- which predicts also $v_2\lesssim$~0.035 (the maximum value obtained when 
the produced strongly interacting matter has vanishing viscosity) 
at 14~TeV. Also percolation models~\cite{Cunqueiro:2008uu} predict integrated $v_2$ of the order of 2\%--3\%.
Although small, such $v_2$ values are comparable to those measured in nucleus-nucleus reactions at c.m. energies 
$\sqrtsnn \approx$~5~GeV at the BNL Alternating Gradient Synchrotron (AGS)~\cite{Barrette:1996rs} or 
$\sqrtsnn \approx$~17~GeV at the CERN Super Proton Synchrotron (SPS)~\cite{Alt:2003ab} and thus could be in principle measurable.

The experimental measurement of $|v_2|$ signals of maximum magnitude $\mathcal{O}(3\%)$, in \pp\ 
collisions at the LHC will be certainly challenging. On the one hand, one will have to deal with all standard 
``non-flow'' effects -- i.e. azimuthal correlations not associated with parton collective rescattering -- generated by 
jets, resonance decays, Hanbury-Brown Twiss (HBT) short-range correlations, or (depending on the technique 
used to determine $v_2$) momentum conservation~\cite{Voloshin:2008dg}. In particular, the large minijet 
production cross section at the LHC generates already strong (back-to-back) azimuthal anisotropies~\cite{Lokhtin1,Lokhtin2}. 
In addition, part of the azimuthal correlations can come, not from final-state rescatterings but 
from parton correlations in the {\it initial} state~\cite{Boreskov:2008uy,Kopeliovich:2008nx}.
On the other hand, the relatively small magnitude of the \pp\ eccentricity and overlap area 
will result in large fluctuations of the $v_2$ parameter even for a same given centrality.
Likewise, event-by-event fluctuations on the produced particle multiplicities will also 
complicate their correlation to the measured $v_2$ values, as commonly presented in plots like that in 
Fig.~\ref{fig:v2pp_vs_cent} (right). 
In any case, there exist detailed methods of elliptic flow analyses developed for nuclear collisions
~\cite{Ollitrault:1992bk,Sorge:1998mk,Voloshin:1994mz,Voloshin1,Voloshin:1999gs,yang,Eyyubova} 
that can be tested and adapted to \pp\ collisions at the LHC. We leave for a future paper the detailed 
study of the experimental feasibility of the measurement. As aforementioned, previous studies in 
heavy ion collisions at lower energies have clearly demonstrated the possibility to measure $v_2$ signals 
with absolute magnitudes of order 1\%--3\% and the ALICE~\cite{tdr_alice}, ATLAS~\cite{Steinberg:2007nm} 
and CMS~\cite{tdr_cms} experiments have proven detector capabilities to carry out such measurements. 

\section{Summary and conclusions}
\label{sec:summ}

We have studied the possibility to observe a collective expansion signal -- in the form of an azimuthal 
anisotropy of particle production with respect to the reaction plane -- due to multi-parton interactions in 
proton-proton collisions at the LHC. Our working assumption has been that any possible azimuthal 
aniso-tropy due to collective flow in \pp\ collisions should follow the same eccentricity, overlap-area and 
particle multiplicity dependences observed in the strongly interacting matter formed in high-energy heavy ion collisions. 
Using a simple eikonal model for multiparton scatterings, we have tested various proton density 
distributions proposed in the literature and obtained the corresponding eccentricities, transverse 
areas and hadron multiplicities as a function of the impact parameter for \pp\ collisions at $\sqrts$~=~14~TeV. 
The transverse overlap area $A_\perp$ and the final particle multiplicity at midrapidity $dN/dy$ are about 
two orders of magnitude smaller in \pp\ compared to \AuAu\ collisions. Since the elliptic flow 
roughly depends on the normalised $(dN/dy)/A_\perp$ ratio, we thus expect any elliptic flow in the 
\pp\ systems to be mostly driven by the eccentricity (if any) of the system produced in the collision.\\

Our first finding is that, popular proton matter profiles such as Gaussian or double-Gaussian result in
vanishing \pp\ eccentricities and, thus, cannot generate any final-state elliptic flow within our approach. 
Unphysical sharp edge (hard-sphere) profiles result in integrated elliptic flow parameters, $v_2 \approx$~10\%, 
almost twice larger than found in \AaAa\ at RHIC. More realistic Fermi-Dirac distributions 
with a diffuse proton edge, yield maximum $v_2$ values of the order of 1.5\%. If the Fermi 
density is too dilute, as in our considered Fermi-II case that takes into account an effective growth 
of the proton size due to the transverse spread of its partons at high energies, the generated $v_2$ 
will be virtually null.
Lastly, exponential proton profiles that reproduce the proton charge form-factor (i.e. the spatial 
distribution of its valence quarks) would result in negative, i.e. out-of-plane, integrated elliptic flows with minimum
values of the order of $-$3\%.

All in all, our work demonstrates that the study of hadron anisotropies with respect to the reaction plane in \pp\ collisions 
at LHC energies, can provide important information on the proton shape and structure at moderate virtualities $\mathcal{O}$(0.7 GeV). 
Although previous analyses with heavy ions at much lower energies have indeed measured integrated $v_2$ 
of a few percents, the experimental extraction of such a signal in \pp\ collisions will be challenging given 
the expected large non-flow azimuthal correlations that can mask the signal. Despite these difficulties, the absence 
or presence of elliptic flow in the data and its dependence on the ``centrality'' of the collision, will nonetheless 
put strong constraints on the density profile of the proton at very high energies.

\section*{Acknowledgments}

We thank Mark Strikman for valuable discussions and comments on a previous version of 
the paper. This research was supported by Russian Foundation for Basic Research 
(grants No 08-02-91001 and No 08-02-92496), Grant of President of Russian 
Federation No 107.2008.2, Dynasty Foundation and Russian Ministry for 
Education and Science (contracts 01.164.1-2.NB05 and 02.740.11.0244).
D.d'E. acknowledges support by the 7th EU Framework Programme (contract FP7-ERG-2008-235071). 

\section*{Appendix I.}
\label{sec:app}
\begin{figure*}[htpb]
\centering
\mbox{
\subfigure{\epsfig{figure=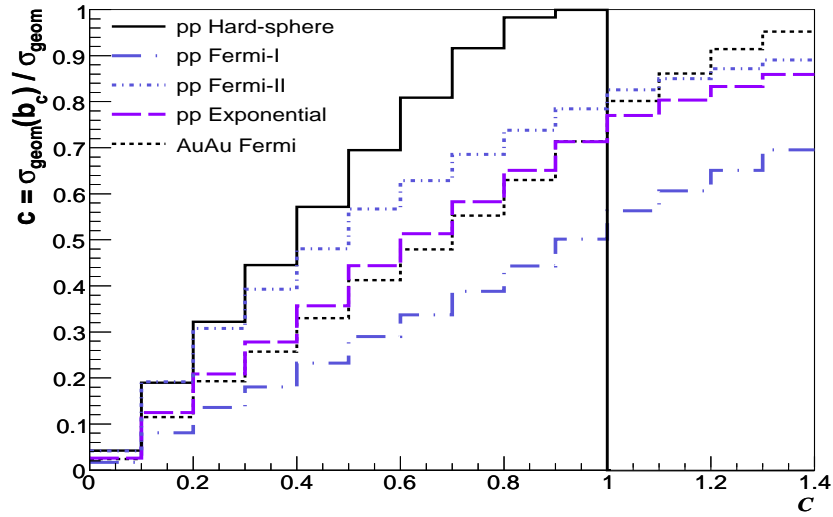,height=7.cm,width=11.cm}}
} 
\caption{Comparison of the exact centrality $c$, Eq.~(\ref{eq:a9_1}), to the centrality $\Cent = b^2/(4R^2)$ 
used in the present work.
\label{fig:c_C}}
\end{figure*}

We recall here the basic transformations from impact parameter to centrality for a
general two-dimensional probability density $F(b)$ which depends on the impact 
parameter vector $\bar b$, 
\begin{eqnarray}
\label{eq:a1} {d^2P\over d^2\vec{b} }={d^2P\over bd b d\varphi}= {1\over2\pi}{dP\over bdb}=F(b)\,,\;\;\; \nonumber \\
\mbox{ with normalisation }
\int d^2b{d^2P\over d^2\vec{b} }= \int db (2\pi b F(b))=1.\nonumber
\end{eqnarray}
The corresponding normalised one-dimensional density for the dimensionless impact parameter $b'=b/(2R)$,
which we have used in this work to easily compare the results obtained with proton densities with 
different radii $R$, is
\begin{equation}
\label{eq:a3} {dP\over db }= 2\pi b F(b)\;\; \to \;\; {dP\over db' }= 4\pi R b F(b).\nonumber
\end{equation}
For our alternative centrality parameter defined as $\Cent= b^2/(4R^2)$, i.e. $db=2R^2/b\;dC$, we have
\begin{eqnarray}
\label{eq:a5} {dP\over dC } = {2R^2\over b }{dP\over db } = {2R^2\over b } 2\pi b F(b)= 4\pi R^2 F(b)\;, \nonumber \\
\mbox{ normalised to }
\int dC{dP\over dC } =  \int {b\over 2R^2}db ~{2R^2\over b }{dP\over db }= \int db {dP\over db } = 1.\nonumber
\end{eqnarray}
Such a definition of $\Cent$ is convenient as it corresponds to a given fraction of the inelastic cross section.
Indeed, for two colliding black disks with radii $R_1$ and  $R_2$ the fraction of the inelastic cross section is
\begin{equation}
\label{eq:a8} {\Delta \sigma_{inel}\over \sigma_{inel} }=
{2\pi b\Delta b\over \pi (R_1+R_2)^2 }= {\Delta b^2\over (R_1+R_2)^2 },~~~b \leq (R_1+R_2)\;,\nonumber
\end{equation}
and thus for two identical disks $R=R_1=R_2$, we have
 \begin{equation}
{\Delta \sigma_{inel}\over \sigma_{inel} }= \Delta \left({ b^2\over 4R^2 }\right)=\Delta C.\nonumber
\label{eq:a9} 
\end{equation}
The exact definition of centrality $c$ as a fraction of the inelastic cross section is equal to~\cite{Vogt}
 \begin{equation}
c={\sigma_{geom}(b_c) \over \sigma_{geom} }= {2\pi\int_0^{b_c} bdb(1-e^{-\sigma_{gg}T_{pp}(b)}) 
\over \sigma_{pp}^{inel} }.\nonumber
\label{eq:a9_1} 
\end{equation}
In Fig.~\ref{fig:c_C} we compare the exact centrality $c$ to the centrality $\Cent$ used in the present work.
\noindent
Let us consider, as an example of $b$-to-$\Cent$ transformation, the case of the probability of inelastic parton scatterings, 
presented in Section~\ref{sec:eikonal}. Such a distribution, which as a function of impact parameter and centrality reads:
\begin{equation}
{dP_{gg}^{inel}\over db }= 2\pi bF(b)= {2\pi b(1-e^{-\sigma_{gg} ~T_{pp}(b)}) \over \int d^2 \vec{b}(1-e^{-\sigma_{gg} ~T_{pp}(b)})}\;,\;\;\mbox{ and }\nonumber
\label{eq:a10} 
\end{equation}
\begin{equation}
{dP_{gg}^{inel}\over dC} = 4\pi R^2 F(b)= {4\pi R^2(1-e^{-\sigma_{gg} ~T_{pp}(b)}) \over \int d^2 \vec{b}(1-e^{-\sigma_{gg} ~T_{pp}(b)})}, ~~b=2R \sqrt{C}\nonumber
\label{eq:a11} 
\end{equation}
is used in our calculations to determine the multiplicity density
\begin{equation}
\frac{d^3N}{dyd^2\vec{b}}(b)=  \frac{dN_{\mbox{\tiny{\it MB}}}}{dy}
 \;\frac{{ \; N_{coll,gg}(b) \; {d^2P_{gg}^{inel}\over d^2\vec{b}}(b)}}
{{\int d^2 \vec{b} \; N_{coll,gg}(b) \; {d^2P_{gg}^{inel}\over d^2\vec{b}}(b)}}\;.\nonumber
\label{eq:dNdyd2b} 
\end{equation}
The multiplicity for different centrality bins is:
\begin{equation}
{dN \over dy}(C_k) = \int^{C_k+\Delta C}_{C_k} dC {d^2N \over dy\,dC}(C).\nonumber
\label{eq:dNdy_vs_b2} 
\end{equation}
The dimensionless multiplicity and probability in a given impact-parameter and centrality bin are
\begin{equation}
N(b)= {dN\over db}\Delta b\;, \mbox { ~~~~~~or ~~~~~} N(C)={dN\over dC}\Delta C,\nonumber
\label{eq:centclassa} 
\end{equation}

\begin{equation}
P_{gg}^{inel}(b)={dP_{gg}^{inel}\over db}\Delta b\nonumber
, \mbox { ~~~~~~or ~~~~~}
P_{gg}^{inel}(C)={dP_{gg}^{inel}\over dC}\Delta C.\nonumber
\label{eq:centclassb} 
\end{equation}



\end{document}